\RequirePackage{fix-cm}
\documentclass[twocolumn,epjc3]{svjour3}
\RequirePackage{graphicx}
\RequirePackage{amsmath}
\RequirePackage{amssymb}
\journalname{Eur. Phys. J. C}

\begin{document}

\title{
One-dimensional soliton system of gauged kink and Q-ball 
}
\author{A.Yu.~Loginov 
        \and
        V.V.~Gauzshtein} 

\institute{Tomsk Polytechnic University, 634050 Tomsk, Russia \label{addr1}}
\date{Received: date / Accepted: date}
\maketitle

\begin{abstract}
In the present  paper,  we  consider  a  $\left(1 + 1\right)$-dimensional gauge
model  consisting  of  two  complex  scalar  fields interacting with each other
through an  Abelian gauge field. When the model's gauge  coupling constants are
set equal to zero, the  model  possesses  non-gauged  Q-ball and kink solutions
that do not interact with each other. It is shown that at nonzero gauge coupling
constants,  the  model  possesses  the  soliton solution describing  the system
consisting of interacting Q-ball and  kink  components.   The  kink and  Q-ball
components of the kink-Q-ball system  have  opposite electric  charges,  so the
total  electric  charge of the kink-Q-ball system  vanishes.  Properties of the
kink-Q-ball system are researched analytically and numerically. In  particular,
it was found that the  kink-Q-ball  system  possesses  a nonzero electric field
 and is unstable with respect to  small  perturbations of fields.
\keywords{kink \and Q-ball \and Noether charge \and gauge field}
\PACS{11.10.Lm \and  11.27.+d}
\end{abstract}

\section{Introduction} \label{sec:I}

It is known  that in the case of Maxwell  electrodynamics,  any one-dimensional
or two-dimensional field configuration with a nonzero electric charge possesses
infinite  energy.
The reason is simple: at large  distances,  the  electric field of such a field
configuration does  not  depend  on  coordinate in the one-dimensional case and
behaves as $r^{-1}$ in the two-dimensional case,  so the energy of the electric
field diverges linearly  in the one-dimensional case and logarithmically in the
two-dimensional case.
That is  why  there  are  no  electrically  charged  solitons  in  one  and two
dimensions; such   solitons    appear    only    in    three  dimensions  (e.g.
three-dimensional   electrically   charged   dyon  \cite{julia_zee}  or  Q-ball
\cite{klee,anag,ardoz_2009,gulamov_2014}).
It should be noted, however, that electrically charged two-dimensional vortices
exist   in   both   the    Chern-Simons   \cite{hong,jw1,jw2,bazeia_1991,ghosh}
and the Maxwell-Chern-Simons \cite{paul,khare_rao_227,khare_255,loginov_plb_784}
gauge models.
Furthermore, it  was  shown  in  \cite{dsantos,losano} that  Chern-Simons gauge
models also possess one-dimensional domain walls.
The domain  walls  have  finite linear  densities of magnetic flux and electric
charge, so there is a linear momentum flow along the domain walls.

Nevertheless,  even  in  the  case  of  Maxwell  gauge  field models, there are
electrically neutral  low-dimensional  soliton systems  with a nonzero electric
field in their interior areas.
In particular, the one-dimensional  soliton  system  consisting of electrically
charged   Q-ball   and   anti-Q-ball   components   has   been   considered  in
\cite{loginov_prd_99} and  the  two-dimensional  soliton  systems consisting of
vortex and Q-ball components interacting  through   an   Abelian   gauge  field
has been described in \cite{loginov_plb_777}.

In the present paper, we research the one-dimensional soliton system consisting
of Q-ball  and  kink  components,  possessing opposite electric charges, so the
system with  a  nonzero  electric  field  is  electrically  neutral as a whole.
Properties of this  kink-Q-ball  system  are  investigated  by  analytical  and
numerical methods.
In particular, we found  that in  contrast  to  the  non-gauged one-dimensional
Q-ball, the kink-Q-ball system does not go into the thin-wall regime.

There is an  interesting  problem  concerning  the stability of the kink-Q-ball
system with respect to small perturbations of fields.
Recall that the  Abelian  Higgs  model  possesses  an electrically neutral kink
solution \cite{bochkarev_mpl_2,grigoriev}.
Formally, this gauged kink solution  is  the  usual  kink of a self-interacting
real scalar field up to gauge transformations. 
However, properties  of  these  two kink solutions differ considerably, because
the classical vacua of the corresponding field models have a different topology.
While the real kink is  topologically  stable, the  gauged kink has exactly one
unstable mode.
From  the  topological  point  of  view,  the  gauged  kink  lies  between  two
topologically distinct vacua of the Abelian Higgs  model,  so it is a sphaleron
\cite{bochkarev_mpl_2,grigoriev}.
Note, however, that  the  gauged  kink  is  a static field configuration modulo
gauge transformation, whereas the kink-Q-ball system will depend on time in any
gauge.
Due to this fact, the kink-Q-ball system cannot be  a sphaleron, so its classic
stability requires separate consideration.

The paper has the following  structure.
In Sec.~\ref{sec:II}, we describe briefly  the  Lagrangian, the symmetries, the
field equations, and the energy-momentum tensor of the Abelian gauge model under
consideration.
In Sec.~\ref{sec:III}, we research properties of the kink-Q-ball system.
Using the  Hamiltonian  formalism  and  the  Lagrange  multipliers  method,  we
establish the time dependence of the soliton system's fields.
The important differential relation for the kink-Q-ball solution is derived and
the system of nonlinear differential equations for ansatz functions is obtained.
Then we establish some general properties of the kink-Q-ball system.
In particular,  we  research  asymptotic  behavior  of  the  system's fields at
small  and   large  distances,  establish  some  important  properties  of  the
electromagnetic  potential, and  derive  the  virial  relation  for the soliton
system.
In Sec.~\ref{sec:IV}, we study properties of the kink-Q-ball system in the three
extreme regimes, namely,  in  the thick-wall regime and in the regimes of small
and large gauge coupling constants.
We also establish basic properties of the plane-wave field configuration of the
model. 
In Sec.~\ref{sec:V},  we  present  and  discuss the numerical results obtained.
They include the dependences  of  the  energy  of the kink-Q-ball system on its
phase frequency and Noether charge, along with numerical results for the ansatz
functions, the energy  density, the  electric  charge density, and the electric
field strength.

Throughout the paper, we use the natural units $\hbar = c = 1$.

\section{The gauge model}                                        \label{sec:II}

The Lagrangian density  of the $\left(1+1\right)$-dimensional gauge model under
consideration has the form
\begin{eqnarray}
\mathcal{L} &=&-\frac{1}{4}F_{\mu \nu }F^{\mu \nu }
+ \left( D_{\mu }\phi \right) ^{\ast }D^{\mu }\phi  - V\left( \left\vert \phi
\right\vert \right) \nonumber \\
&&+\left( D_{\mu }\chi \right) ^{\ast }D^{\mu }\chi - U\left( \left\vert \chi
\right\vert \right),                                               \label{II:1}
\end{eqnarray}
where $F_{\mu \nu}=\partial_{\mu}A_{\nu}-\partial_{\nu}A_{\mu}$ is the strength
of the  Abelian  gauge  field  and  $\phi$, $\chi$  are  complex  scalar fields
minimally  interacting  with  the Abelian gauge  field  through  the  covariant
derivatives:
\begin{equation}
D_{\mu }\phi  = \partial_{\mu }\phi + i e A_{\mu }\phi, \quad
D_{\mu }\chi  = \partial_{\mu }\chi + i q A_{\mu }\chi.            \label{II:2}
\end{equation}
The self-interaction potentials of the scalar fields have the form
\begin{eqnarray}
V\left(\left\vert \phi \right\vert \right) & = & \frac{\lambda }{2}\left(
\left\vert \phi \right\vert^{2} - \eta^{2}\right)^{2},            \label{II:3a}
\\
U\left(\left\vert \chi \right\vert \right) & = & m_{\chi}^{2}\left\vert \chi
\right\vert^{2}-\frac{g_{\chi}}{2}\left\vert \chi \right\vert^{4}+
\frac{h_{\chi }}{3}\left\vert \chi \right\vert^{6}.               \label{II:3b}
\end{eqnarray}
Let us  suppose that the  self-interaction  potential  $U\left( \left\vert \chi
\right\vert \right)$  has  a  global zero minimum at  $\chi = 0$ and admits the
existence of usual non-gauged Q-balls.
Then the parameters $g_{\chi}$  and  $h_{\chi}$  are  positive  and satisfy the
condition $3g_{\chi }^{2}<16h_{\chi }m_{\chi }^{2}$.
Unlike the sixth-order  potential $U\left(\left\vert \chi \right\vert \right)$,
the fourth-order  potential $V\left(\left\vert \phi \right\vert\right)$ reaches
a zero minimum on the circle $\left\vert \phi \right\vert = \eta$.
The  potential $V\left(\left\vert \phi \right\vert\right)$ allows the existence
of the complex non-gauged kink solution
\begin{equation}
\phi _{\mathrm{k}}\left( x\right) = \eta \tanh\left(\frac{m_{\phi}x}{2}\right)
\exp \left(- i \delta \right),                                     \label{II:4}
\end{equation}
where  $m_{\phi }  =  \sqrt{2 \lambda } \eta$   is   the   mass  of  the scalar
$\phi$-particle and $\delta$ is an arbitrary phase.

Besides  the  local  gauge  transformations:
\begin{subequations} \label{II:4a}
\begin{eqnarray}
\phi \left( x\right)  &\rightarrow &\phi ^{\prime }\left( x\right) =\exp
\left(- i e \Lambda \left( x\right) \right) \phi\left( x\right) ,  \\
\chi \left( x\right)  &\rightarrow &\chi ^{\prime }\left( x\right) =\exp
\left(-  iq\Lambda \left( x\right) \right) \chi\left( x\right) ,  \\
A_{\mu }\left( x\right)  &\rightarrow &A_{\mu }^{\prime }\left( x\right)
= A_{\mu }\left( x\right) + \partial _{\mu }\Lambda \left( x\right),
\end{eqnarray}
\end{subequations}
the Lagrangian (\ref{II:1}) is  also  invariant under the two {\it independent}
global gauge  transformations:
\begin{subequations} \label{II:5}
\begin{eqnarray}
\phi \left( x\right)  &\rightarrow &\phi ^{\prime }\left( x\right) = \exp
\left(-i \alpha \right) \phi \left( x\right),
\\
\chi \left( x\right)  &\rightarrow &\chi ^{\prime }\left( x\right) = \exp
\left(-i \beta \right) \chi \left( x\right).
\end{eqnarray}
\end{subequations}
As a consequence, we have the two Noether currents:
\begin{subequations} \label{II:6}
\begin{eqnarray}
j_{\phi }^{\mu } &=&i\left( \phi ^{\ast }D^{\mu }\phi - \left(D^{\mu }\phi
\right) ^{\ast }\phi \right),                                     \label{II:6a}
\\
j_{\chi }^{\mu } &=&i\left( \chi ^{\ast }D^{\mu }\chi -\left( D^{\mu }\chi
\right) ^{\ast }\chi \right),                                     \label{II:6b}
\end{eqnarray}
\end{subequations}
and the two separately conserved Noether charges: $Q_{\phi}=\int j_{\phi}^{0}dx$
and $Q_{\chi}=\int j_{\chi}^{0}dx$.
Note also that in  addition  to  the  local  and  global gauge transformations,
the Lagrangian (\ref{II:1}) is  invariant  under the discrete $C$, $P$, and $T$
transformations.

The field equations of the model are written as
\begin{eqnarray}
\partial_{\mu }F^{\mu \nu } &  =  &j^{\nu},                       \label{II:7a}
\\
D_{\mu }D^{\mu }\phi +\frac{\partial V}{\partial \left\vert \phi \right\vert
}\frac{\phi }{2\left\vert \phi \right\vert } &=&0,
                                                                  \label{II:7b}
\\
D_{\mu }D^{\mu }\chi +\frac{\partial U}{\partial \left\vert \chi \right\vert
}\frac{\chi }{2\left\vert \chi \right\vert } &=&0,
                                                                  \label{II:7c}
\end{eqnarray}
where  the electromagnetic current $j^{\nu}$ is written in terms of two Noether
currents (\ref{II:6})
\begin{equation}
j^{\nu } = ej_{\phi }^{\nu } + qj_{\chi }^{\nu }.                  \label{II:8}
\end{equation}
From Eq.~(\ref{II:8})  it  follows  that  the electric charges $e Q_{\phi}$ and
$q Q_{\chi}$ of  the  complex  scalar  fields  $\phi$  and $\chi$ are conserved
separately.
This fact is  a  consequence  of  the  neutrality  of  the  Abelian gauge field
$A_{\mu}$.

The symmetric energy-momentum  tensor  of  the  model can be obtained  by using
the well-known formula $T_{\mu \nu} = 2\partial \mathcal{L}/\partial g^{\mu\nu}
-g_{\mu \nu} \mathcal{L}$:
\begin{eqnarray}
T_{\mu \nu }& =&-F_{\mu \lambda }F_{\nu }^{\;\lambda }+\frac{1}{4}g_{\mu \nu
}F_{\lambda \rho }F^{\lambda \rho } \nonumber \\
& &+\left( D_{\mu }\phi \right) ^{\ast }D_{\nu }\phi +\left( D_{\nu }\phi
\right) ^{\ast }D_{\mu }\phi \nonumber \\
& &+\left( D_{\mu }\chi \right) ^{\ast }D_{\nu }\chi +\left( D_{\nu }\chi
\right) ^{\ast }D_{\mu }\chi  \nonumber \\
& &-g_{\mu \nu }\left[ \left( D_{\mu }\phi \right) ^{\ast }D^{\mu }\phi
+\left( D_{\mu }\chi \right) ^{\ast }D^{\mu }\chi \right. \nonumber \\
& &\left. -V\left( \left\vert \phi \right\vert \right) -U\left( \left\vert
\chi \right\vert \right) \right].                                  \label{II:9}
\end{eqnarray}
Thus,  we  have  the  following  expression for the energy density of the model
\begin{eqnarray}
T_{00} = \mathcal{E} &=&\frac{1}{2}E_{x}^{2} +
\left( D_{t}\phi \right)^{\ast }D_{t}\phi
+\left( D_{x}\phi \right) ^{\ast }D_{x}\phi \nonumber \\
&&+\left( D_{t}\chi \right) ^{\ast }D_{t}\chi + \left( D_{x}\chi \right)
^{\ast }D_{x}\chi \nonumber \\
&&+V\left( \left\vert \phi \right\vert \right) + U\left( \left\vert \chi
\right\vert \right),                                              \label{II:10}
\end{eqnarray}
where $E_{x} = F_{01} = \partial_{t} A_{1}-\partial_{x} A_{0}$  is the electric
field strength.

\section{The kink-Q-ball system and its properties}             \label{sec:III}

It is known \cite{lee,fried}  that any nontopological soliton, in particular, a
Q-ball,  is  an  extremum  of  an  energy  functional  at  a fixed value of the
corresponding Noether charge.
Using  this  basic  property  of  a  Q-ball and  taking  into  account that the
self-interaction potential $U\left( \left\vert \chi \right\vert \right)$ admits
the existence of Q-balls formed from the complex  scalar field $\chi$, we shall
search for a soliton  solution  of  model (\ref{II:1})  that  is an extremum of
the energy functional $E = \int\nolimits_{-\infty}^{\infty}\mathcal{E} dx$ at a
fixed value of the  Noether charge $Q_{\chi} = \int\nolimits_{-\infty}^{\infty}
j_{\chi}^{0}dx$.
According  to  the  method  of  Lagrange  multipliers,  such  a  solution is an
unconditional extremum of the functional
\begin{equation}
F = \int\limits_{-\infty}^{\infty} \mathcal{E} dx -
\omega \int\nolimits_{-\infty}^{\infty} j_{\chi}^{0}dx =
E - \omega Q_{\chi},                                              \label{III:1}
\end{equation}
where $\omega$ is the Lagrange multiplier.
Let us determine the time dependence of the soliton solution.
To do  this, we shall  use  Eq.~(\ref{III:1})  and  the  Hamiltonian formalism.
In the  axial  gauge  $A_{x} = A^{1} = 0$,  the  Hamiltonian  density  of model
(\ref{II:1}) is written as
\begin{eqnarray}
\mathcal{H}& = &\pi _{\phi }\partial _{t}\phi +\pi _{\phi ^{\ast }}
\partial_{t}\phi ^{\ast }+\pi _{\chi }\partial_{t}\chi +
\pi _{\chi ^{\ast}}\partial _{t}\chi ^{\ast } - \mathcal{L}
\nonumber \\
& = & -\frac{1}{2}\left( \partial _{x}A_{0}\right) ^{2}+\pi_{\phi }\pi_{\phi
^{\ast }}+\pi _{\chi }\pi _{\chi ^{\ast }}
\nonumber \\
& &  + \partial _{x}\phi ^{\ast }\partial _{x}\phi + \partial _{x}\chi^{\ast
}\partial _{x}\chi
\nonumber \\
& & + ieA_{0}\left\{ \phi ^{\ast }\pi _{\phi ^{\ast }} - \phi \pi _{\phi
}\right\} + iqA_{0}\left\{ \chi ^{\ast }\pi _{\chi ^{\ast }}-\chi \pi _{\chi
}\right\}
\nonumber \\
& & +V\left( \left\vert \phi \right\vert \right) +U\left( \left\vert \chi
\right\vert \right).                                              \label{III:2}
\end{eqnarray}
We see  that  in  the  adopted  gauge,  the  model is described in terms of the
eight canonically conjugated fields:  $\phi$,  $\pi _{\phi } = \left( D_{0}\phi
\right)^{\ast }$,   $\phi^{\ast}$,   $\pi_{\phi^{\ast }}  = D_{0}\phi$, $\chi$,
$\pi_{\chi} = \left(D_{0}\chi \right)^{\ast }$, $\chi^{\ast}$, and $\pi_{\chi^{
\ast}} = D_{0}\chi$,  while  the  time  component $A_{0}$ of the gauge field is
determined  in  terms  of  the  canonically  conjugated  fields  by Gauss's law
\begin{equation}
\partial _{x}^{2}A_{0}+ie\left\{ \phi ^{\ast }\pi _{\phi ^{\ast }}-\phi \pi
_{\phi }\right\} + i q \left\{ \chi ^{\ast } \pi_{\chi ^{\ast }} - \chi \pi
_{\chi }\right\} = 0,                                             \label{III:3}
\end{equation}
and so it is not an independent dynamic field.
Although  energy  density  (\ref{II:10})  is  not equal to Hamiltonian  density
(\ref{III:2}):
\begin{eqnarray}
\mathcal{H-E} &\mathcal{=}&\mathcal{-}\left( \partial _{x}A_{0}\right)
^{2}+ieA_{0}\left\{ \phi ^{\ast }\pi _{\phi ^{\ast }}-\phi \pi _{\phi
}\right\} \nonumber \\
&&+iqA_{0}\left\{ \chi ^{\ast }\pi _{\chi ^{\ast }} - \chi \pi _{\chi
}\right\},                                                        \label{III:4}
\end{eqnarray}
the  integral  of  Eq.~(\ref{III:4}) over  the  one-dimensional  space vanishes
provided  that  field  configurations of the  model possess  finite  energy and
satisfy Gauss's law (\ref{III:3}).

In the adopted axial  gauge  $A_{x}  =  0$,  field  equations (\ref{II:7b}) and
(\ref{II:7c}) can be recast in the Hamiltonian form:
\begin{eqnarray}
\partial _{t}\phi  &=&\frac{\delta H}{\delta \pi _{\phi }}=\frac{\delta E}
{\delta \pi _{\phi }},\quad\partial _{t}\pi _{\phi }=-\frac{\delta H}{\delta
\phi } = -\frac{\delta E}{\delta \phi },                         \label{III:5a}
\\
\partial _{t}\chi  &=&\frac{\delta H}{\delta \pi _{\chi }}=\frac{\delta E}
{\delta \pi _{\chi }},\quad\partial _{t}\pi _{\chi }=-\frac{\delta H}{\delta
\chi } = -\frac{\delta E}{\delta \chi},                          \label{III:5b}
\end{eqnarray}
where we use the  relation $E = \int\nolimits_{-\infty}^{\infty} \mathcal{E} dx
= H = \int\nolimits_{-\infty}^{\infty} \mathcal{H} dx$.
On the other hand, the first variation  of functional (\ref{III:1}) vanishes on
the soliton solution:
\begin{equation}
\delta F = \delta E - \omega \delta Q_{\chi } = 0,                \label{III:6}
\end{equation}
where the  first  variation  of  the  Noether charge $Q_{\chi}$ is expressed in
terms of the canonically conjugated fields as follows:
\begin{equation}
\delta Q_{\chi } = -i\int \left(\pi _{\chi}\delta \chi+\chi \delta \pi_{\chi}
-\text{c.c.}\right) dx.                                           \label{III:7}
\end{equation}
Combining  Eqs.~(\ref{III:5a}) -- (\ref{III:7}),  we  find  that in the adopted
gauge, only the time derivatives  of  the canonically conjugated fields $\chi$,
$\pi_{\chi}$, $\chi^{\ast}$, and $\pi_{\chi^{\ast}}$  are  different from zero:
\begin{eqnarray}
\partial _{t}\chi  &=&\frac{\delta H}{\delta \pi _{\chi }}=\omega \frac{
\delta Q_{\chi }}{\delta \pi _{\chi }}=-i\omega \chi ,           \label{III:8a}
\\
\partial _{t}\pi _{\chi } &=&-\frac{\delta H}{\delta \chi }=-\omega \frac{
\delta Q_{\chi }}{\delta \chi }=i\omega \pi _{\chi },            \label{III:8b}
\\
\partial _{t}\chi ^{\ast } &=&\frac{\delta H}{\delta \pi _{\chi ^{\ast }}}
=\omega \frac{\delta Q_{\chi }}{\delta \pi _{\chi ^{\ast }}}=i\omega \chi
^{\ast },                                                        \label{III:8c}
\\
\partial _{t}\pi _{\chi ^{\ast }} &=&-\frac{\delta H}{\delta \chi ^{\ast }}
=-\omega \frac{\delta Q_{\chi }}{\delta \chi ^{\ast }}=-i\omega \pi _{\chi
^{\ast }},                                                       \label{III:8d}
\end{eqnarray}
while the time derivatives  of $\phi$, $\pi_{\phi}$, $\phi^{\ast}$, and  $\pi_{
\phi^{\ast}}$  are equal to zero.
Recalling that $\pi_{\chi } = \left( D_{0}\chi \right) ^{\ast } = \partial _{t}
\chi^{\ast} -i q A_{0} \chi^{\ast}$ and taking into account Eqs.~(\ref{III:8b})
and (\ref{III:8c}),  we  conclude  that  the  time  derivative  of $A_{0}$ also
vanishes.
It follows  that  only  the  scalar  field  $\chi$  of  the soliton  system has
nontrivial time dependence:
\begin{subequations} \label{III:9}
\begin{eqnarray}
\phi \left( x,t\right)  & = & f\left( x \right),                 \label{III:9a}
\\
\chi \left( x,t\right)  & = & s\left( x \right) \exp\left(-i\omega t\right),
                                                                 \label{III:9b}
\\
A_{\mu}\left(x,t\right) & = & \left(a_{0}\left( x \right), 0 \right).
                                                                 \label{III:9c}
\end{eqnarray}
\end{subequations}

Let us return to Eq.~(\ref{III:6}).
This equation holds for arbitrary variations of fields on the soliton solution,
including those that transfer the soliton solution to  an infinitesimally close
one.
It follows that  the  energy  of  the  soliton  system  satisfies the important
relation
\begin{equation}
\frac{dE}{dQ_{\chi}} = \omega,                                   \label{III:10}
\end{equation}
where  the  Lagrange multiplier $\omega$ is some function of the Noether charge
$Q_{\chi}$.
Since the energy $E$  and  the  Noether charge $Q_{\chi}$ of the soliton system
are  gauge-invariant, relation  (\ref{III:10}) is also gauge-invariant.
Like the case of non-gauged nontopological  solitons  \cite{fried,coleman,lee},
relation (\ref{III:10}) determines  basic  properties of the gauged kink-Q-ball
system.

In Eqs.~(\ref{III:9}), functions  $f\left(x\right)$  and  $s\left(x\right)$ are
assumed to be some complex functions of the real argument $x$.
Substituting   Eqs.~(\ref{III:9})   into   field   equations  (\ref{II:7a})  --
(\ref{II:7c}),  we  can  easily  check  that  the  real  and imaginary parts of
$f\left(x\right)$ satisfy the same differential equation with real coefficients.
Similarly, the real and imaginary parts of $s\left(x\right)$  also  satisfy the
same differential equation with real coefficients.
It follows that the functions $f\left(x\right)$ and  $s\left(x\right)$ have the
form:  $f\left(x\right) = \exp \left( i \alpha \right)\tilde{f}\left(x\right)$,
$s\left( x \right) = \exp \left( i\beta \right) \tilde{s}\left(x\right)$, where
$\tilde{f}\left( x \right)$ and $\tilde{s}\left( x \right)$ are real functions,
whereas $\alpha$ and $\beta$ are constant phases.
However,  these  phases  can  be  cancelled  by  global  gauge  transformations
(\ref{II:5}), so the functions $f\left(x\right)$  and  $s\left(x\right)$ can be
supposed to be real without loss of generality.
The functions $a_{0}\left(x\right)$, $f\left(x\right)$,  and  $s\left(x\right)$
satisfy the system of ordinary nonlinear differential equations:
\begin{equation}
a_{0}^{\prime \prime }(x)-2a_{0}\left( x\right) \left(e^{2}f\left(x\right)^{2}
+q^{2}s\left(x\right)^{2}\right)+2q\omega s\left(x\right)^{2}=0, \label{III:11}
\end{equation}
\begin{equation}
f^{\prime \prime }\left(x\right)+\left(\lambda \eta^{2} +
e^{2}a_{0}\left(x\right)^{2}\right) f\left( x\right) -
\lambda f\left( x\right)^{3}=0,                                  \label{III:12}
\end{equation}
\begin{eqnarray}
& & s^{\prime \prime }\left( x\right) -\left( m_{\chi }^{2}-\left( \omega
-q a_{0}\left( x\right) \right)^{2}\right) s \left( x\right) +
g_{\chi}s\left( x\right)^{3}                                     \label{III:13}
\\
& & -h_{\chi }s\left( x\right)^{5}=0,    \nonumber
\end{eqnarray}
which is  obtained  by  substituting Eqs.~(\ref{III:9})  into  field  equations
(\ref{II:7a}) -- (\ref{II:7c}).

The most important among the local quantities of the kink-Q-ball system are the
electromagnetic current density and the energy density.
Their  expressions  in  terms  of  $a_{0}\left(x\right)$,  $f\left( x \right)$,
and $s\left(x\right)$ are written as
\begin{equation}
j^{\mu }=\left( 2q\omega s^{2}-2a_{0}\left( e^{2}f^{2}+q^{2}s^{2}\right)
,\,0\right),                                                     \label{III:14}
\end{equation}
\begin{eqnarray}
\mathcal{E} & = & \frac{a_{0}^{\prime 2}}{2} + f^{\prime 2} + s^{\prime2}
+ \left( \omega - qa_{0}\right)^{2}s^{2} + e^{2}a_{0}^{2}f^{2}   \label{III:15}
\\
& & + V \left( f \right) + U\left( s \right). \nonumber
\end{eqnarray}
The energy $E=\int\nolimits_{-\infty}^{\infty}\mathcal{E}dx$ of the kink-Q-ball
system must be finite.
Using this fact and Eq.~(\ref{III:15}),  we  obtain  the boundary condition for
$a_{0}\left(x\right)$,  $f\left( x \right)$, and $s\left(x\right)$:
\begin{subequations}                                             \label{III:16}
\begin{eqnarray}
&&a_{0}\left(x\right) \underset{x \rightarrow -\infty}{\longrightarrow} 0,\;
\;\, a_{0}\left(x\right) \underset{x\rightarrow \infty}{\longrightarrow}0,
\label{III:16a} \\
&&f\left(x\right)\underset{x\rightarrow -\infty}{\longrightarrow}\!\!-\eta,
\;\,f\left( x\right) \underset{x\rightarrow \infty }{\longrightarrow} \eta,
\label{III:16b} \\
&&s\left(x\right)    \underset{x \rightarrow -\infty }{\longrightarrow} 0,\;
\;\;\;s\left( x\right) \underset{x\rightarrow \infty }{\longrightarrow} 0.
\label{III:16c}
\end{eqnarray}
\end{subequations}
Note that the finiteness of the electric field's energy $E^{\left( E\right) } =
\int\nolimits_{-\infty }^{\infty} a_{0}^{\prime 2}/2 dx$   leads  to  one  more
boundary condition for $a_{0}\left(x\right)$:
\begin{eqnarray}
& a_{0}^{\prime }\left( x\right) \underset{x\rightarrow - \infty}
{\longrightarrow }0,\; &a_{0}^{\prime }\left( x\right)
\underset{x\rightarrow\infty }{\longrightarrow }0.              \label{III:16d}
\end{eqnarray}
This condition,  however,  is  equivalent  to Eq.~(\ref{III:16a}) provided that
$a_{0}\left(x\right)$ is regular as $x \rightarrow \pm \infty$.

Gauss's law (\ref{III:11}) can  be written as $a_{0}^{\prime \prime} = -j^{0}$,
where $j^{0}$ is electric charge density (\ref{III:14}).
Integrating this equation  over $x \in \left(-\infty, \infty\right)$ and taking
into account  boundary  conditions (\ref{III:16d}),  we conclude that the total
electric  charge  $Q  = \int\nolimits_{-\infty}^{\infty}j^{0}dx$   of  a  field
configuration  with  finite energy vanishes:
\begin{equation}
Q = e Q_{\phi} + q Q_{\chi } = 0,                               \label{III:16e}
\end{equation}
where $Q_{\phi}$   and   $Q_{\phi}$   are   the   Noether  charges  defined  by
Eqs.~(\ref{II:6}).

It  can  easily  be  checked  that  system  (\ref{III:11}) -- (\ref{III:13}) is
invariant under the discrete transformation
\begin{equation}
\omega, a_{0}, f, s \longrightarrow -\omega, -a_{0}, f, s.       \label{III:17}
\end{equation}
This invariance is  a  consequence  of  the  $C$-invariance  of  the Lagrangian
(\ref{II:1}).
Using  Eqs.~(\ref{III:14}), (\ref{III:15}), and  (\ref{III:17}),  we  find  the
behavior of  the energy $E$ and the  Noether  charges $Q_{\phi}$ and $Q_{\chi}$
under the transformation $\omega \rightarrow -\omega$: 
\begin{eqnarray}
E\left( -\omega \right) &=&E\left( \omega \right),              \label{III:18a}
\\
Q_{\phi ,\chi }\left( -\omega \right) &=&-Q_{\phi,\chi}\left(\omega
\right).                                                        \label{III:18b}
\end{eqnarray}
We  see  that  the  energy  of  the  kink-Q-ball  system is an even function of
$\omega$, whereas  the  Noether  charges  $Q_{\phi}$  and  $Q_{\chi}$  are  odd
functions of $\omega$.

The $P$-invariance  of the  Lagrangian  (\ref{II:1}) leads to the invariance of
system (\ref{III:11}) -- (\ref{III:13}) under the space inversion $x\rightarrow
-x$.
Due to the space homogeneity,  the system  (\ref{III:11}) -- (\ref{III:13})  is
also  invariant   under   the   coordinate  shift $x  \rightarrow x  +  x_{0}$.
Furthermore,   due   to   Eqs.~(\ref{II:5}),   the   system  (\ref{III:11})  --
(\ref{III:13}) is invariant under the two independent discrete transformations:
$f \rightarrow -f$ and $s \rightarrow -s$.
These facts and symmetry properties of boundary conditions (\ref{III:16})  lead
to the conclusion that $a_{0}\left(x\right)$  and  $s\left(x \right)$  are even
functions of $x$, while $f\left( x \right)$ is an odd function of $x$.
This is consistent with the fact that the  non-gauged  kink  solution is an odd
function of $x$,  whereas the non-gauged Q-ball solution is an even function of
$x$.

The asymptotic form  of  the  soliton  solution  for  small  $x$ is obtained by
substitution  of  the  power  expansions  for  $a_{0}\left(x\right)$, $f\left(x
\right)$, and $s\left(x\right)$ into  Eqs.~(\ref{III:11}) -- (\ref{III:13}) and
equating the resulting Taylor coefficients to zero.
By acting in this way, we obtain:
\begin{subequations}\label{III:19}
\begin{eqnarray}
a_{0}\left( x\right)  &=&a_{0}+\frac{a_{2}}{2!}x^{2}+O\left( x^{3}\right),
                                                                \label{III:19a}
\\
f_{0}\left( x\right)  &=&f_{1}x+\frac{f_{3}}{3!}x^{3}+O\left( x^{5}\right),
                                                                \label{III:19b}
\\
s_{0}\left( x\right)  &=&s_{0}+\frac{s_{2}}{2!}x^{2}+O\left( x^{3}\right),
                                                                \label{III:19c}
\end{eqnarray}
\end{subequations}
where the next-to-leading coefficients
\begin{subequations}\label{III:20}
\begin{eqnarray}
a_{2} &=&-2qs_{0}^{2}\left( \omega -qa_{0}\right), \label{III:20a} \\
f_{3} &=&-\frac{1}{2}f_{1}\left( m_{\phi }^{2}+2e^{2}a_{0}^{2}\right),  \\
s_{2} &=&s_{0}\left( m_{\chi }^{2}-\left( \omega -qa_{0}\right) ^{2}\right)
-g_{\chi }s_{0}^{3}+h_{\chi }s_{0}^{5}
\end{eqnarray}
\end{subequations}
are expressed in terms  of  the  three  leading  coefficients $a_{0}$, $f_{1}$,
$s_{0}$, and the model's parameters.

For large $\left\vert x\right\vert$, system (\ref{III:11}) -- (\ref{III:13}) is
linearized and we obtain the asymptotic form of the soliton solution satisfying
boundary conditions (\ref{III:16}):
\begin{subequations}\label{III:21}
\begin{eqnarray}
f(x) &\sim &\pm \eta \pm f_{\infty }\exp \left( \mp m_{\phi }x\right),
                                                               \label{III:21aa}
\\
s\left( x\right)  &\sim &s_{\infty }\exp \left( \mp \Delta x \right),
                                                               \label{III:21bb}
\\
a_{0}\left( x\right)  &\sim &a_{\infty }\exp \left( \mp m_{A}x\right)
                                                               \label{III:21cc}
\\
&&-\frac{2q\omega}{4\Delta^{2}-m_{A}^{2}}s_{\infty}^{2}\exp
\left( \mp 2 \Delta x \right), \nonumber
\end{eqnarray}
\end{subequations}
where $m_{\phi}=\sqrt{2\lambda}\eta$, $\Delta = \left(m_{\chi}^{2} - \omega^{2}
\right)^{1/2}$\!, and $m_{A}=\sqrt{2} e \eta$.

Let us discuss the global behavior of the electromagnetic potential $a_{0}\left(
x\right)$.
Since the  total  electric  charge $Q = \int\nolimits_{-\infty}^{+\infty }j^{0}
\left( x\right) dx$ of the kink-Q-ball system  vanishes,  the  electric  charge
density $j^{0}\left( x\right)$  must  vanish  at some  points  of the $x$-axis.
Because of the symmetry  $j^{0}\left(-x\right)  =  j^{0}\left( x\right)$, these
points (nodes  of  $j^{0}\left( x \right)$)  are  symmetric with respect to the
origin $x = 0$.
Next,  according  to  Gauss's  law $a_{0}^{\prime \prime}\left( x\right) = -j^{0}
\left( x\right)$, the second
derivative $a_{0}^{\prime \prime}\left( x \right)$  vanishes  at  the  nodes of
$j^{0}\left( x \right)$.
Thus the nodes  of  $j^{0}\left( x\right)$  are  the  inflection  points of the
electromagnetic potential $a_{0}\left(x\right)$.
From Eq.~(\ref{III:11}) it follows that in an inflection point $x_{\mathrm{i}}$,
the  electromagnetic   potential  $a_{0}\left( x_{\mathrm{i}} \right)$  can  be
expressed in terms of $f\left(x_{\mathrm{i}}\right)$ and $s\left(x_{\mathrm{i}}
\right)$:
\begin{equation}
a_{0}\left(x_{\mathrm{i}}\right) = \frac{\omega qs\left(x_{\mathrm{i}}
\right)^{2}}{e^{2}f\left(x_{\mathrm{i}}\right)^{2}+q^{2}s\left(x_{\mathrm{i}}
\right) ^{2}}.                                                  \label{III:21a}
\end{equation}
Two conclusions follow from Eq.~(\ref{III:21a}).
Firstly, at an inflection point $x_{\mathrm{i}}$,  the  sign of $a_{0}\left(x_{
\mathrm{i}}\right)$ coincides with the sign  of  $\omega$  (we suppose that the
gauge coupling constants are positive by definition): 
\begin{equation}
\text{sign}\left(a_{0}\left(x_{\mathrm{i}}\right)\right) =
\text{sign}\left( \omega \right).                              \label{III:21ab}
\end{equation}
Secondly, at  an  inflection  point  $x_{\mathrm{i}}$, the following inequality
holds:
\begin{equation}
\left\vert a_{0}\left( x_{\mathrm{i}}\right) \right\vert <
\frac{\left\vert\omega \right\vert }{q}.                        \label{III:21b}
\end{equation}
Next, from  Eq.~(\ref{III:14}), we  obtain  the  expression  for  the  electric
charge density at the origin:
\begin{equation}
j^{0}\left( 0\right) = - a_{0}^{\prime \prime}\left( 0\right) =
2 q s \left( 0\right) ^{2} \left( \omega - q a_{0}
\left(0\right) \right),                                         \label{III:21c}
\end{equation}
from which it follows that the sign of the curvature of $a_{0}\left( x \right)$
at $x = 0$ is opposite in  sign  to  $\omega - q a_{0}\left(0\right)$:
\begin{equation}
\text{sign}\left( a_{0}^{\prime \prime}\left( 0\right) \right) =
- \text{sign}\left( \omega -qa_{0}\left(0\right) \right).       \label{III:21d}
\end{equation}
An  elementary   graphical   analysis   made   using   Eqs.~(\ref{III:21a})  --
(\ref{III:21d}) leads  us  to  the  following conclusions about the behavior of
$a_{0}\left( x \right)$:
\begin{equation}
0 < a_{0}\left(\pm x_{\mathrm{i1}}\right) < a_{0}\left( 0\right) <
\frac{\omega }{q}\quad \text{for} \quad \omega > 0,             \label{III:21e}
\end{equation}
and
\begin{equation}
\frac{\omega}{q} < a_{0}\left(0\right) < a_{0}\left(\pm x_{\mathrm{i1}}\right)
<0\quad \text{for} \quad\omega < 0,                             \label{III:21f}
\end{equation}
where $\pm x_{\mathrm{i1}}$  is  the two symmetric inflection points closest to
the origin $x = 0$.
From Eqs.~(\ref{III:21c}), (\ref{III:21e}), and (\ref{III:21f}) it follows that
the sign of the  electric charge density  at  the origin coincides with that of
the phase frequency
\begin{equation}
\text{sign}\left( j^{0}\left( 0\right) \right) =
- \text{sign}\left( a_{0}^{\prime \prime}\left( 0 \right) \right) =
\text{sign}\left(\omega\right).                                 \label{III:21g}
\end{equation}

We can also make conclusions about  the  behavior of $a_{0}\left( x\right)$ for
$\left\vert x \right\vert > x_{\mathrm{i1}}$.
In  particular,  $a_{0}\left( x \right)$   cannot  vanish  at  any  finite $x$.
Indeed, let $x_{\mathrm{n}}$  be  a  conjectural  point  in which $a_{0}\left(x
\right)$ vanishes.
Then from Eq.~(\ref{III:11}) we have the relation
\begin{equation}
a_{0}^{\prime \prime }\left( x_{\mathrm{n}}\right) =-2q\omega s\left( x_{
\mathrm{n}}\right)^{2}.                                         \label{III:21h}
\end{equation}
We see that  the  sign  of  the curvature of $a_{0}\left(x\right)$ at the point
$x_{\mathrm{n}}$ is opposite to the sign of $\omega$.
Let $\omega$ be positive.
Then  from   Eq.~(\ref{III:21h})  it  follows  that  in  some  neighborhood  of
$x_{\mathrm{n}}$, the functions $a_{0}\left(x\right)$ and $a_{0}^{\prime\prime}
\left(x\right)$ are negative.
But according  to  Eq.~(\ref{III:21ab}), there  are  no  inflection  points for
negative $a_{0}\left( x \right)$, so $a_{0}^{\prime\prime}\left( x \right)$ can
never change  the  sign,  $a_{0}\left( x \right)$  decreases  indefinitely, and
boundary condition (\ref{III:16a}) cannot be satisfied.
It follows  that  $a_{0}\left( x \right)$  cannot  vanish  at  any  finite $x$.
The case of negative $\omega$ is treated similarly.
Thus, we  come  to  an important conclusion that  the electromagnetic potential
$a_{0}\left( x \right)$ cannot vanish at any finite $x$, and so the sign of the
electromagnetic potential coincides  with  that of the phase frequency over the
whole range of $x$:
\begin{equation}
\text{sign}\left( a_{0}\left( x\right) \right) = \text{sign}
\left( \omega \right)                                           \label{III:21i}
\end{equation}
for all x.
Of course, this  conclusion  is  valid  only  for adopted gauge (\ref{III:9c}).

Let  $a_{0}\left(x\right)$,  $f\left(x\right)$,  and   $s\left(x\right)$  be  a
solution  of  system  (\ref{III:11})  --  (\ref{III:13}) that  satisfy boundary
conditions (\ref{III:16}).
When we perform the  scale  transformation  $x  \rightarrow  \lambda  x$ of the
argument of the solution, the Lagrangian $L  = \int\nolimits_{-\infty}^{\infty}
\mathcal{L} dx$  becomes  a  simple  function of the scale parameter $\lambda$.
The function  $L \left( \lambda\right)$ must have an extremum at $\lambda = 1$,
so the derivative $dL/d\lambda$  vanishes at this point.
Using this fact, we obtain   the   virial  relation  for  the   soliton system:
\begin{equation}
E^{\left( E\right) }+E^{\left( P\right) }-E^{\left( G\right) } - E^{\left(
T\right) } = 0,                                                  \label{III:22}
\end{equation}
where
\begin{equation}
E^{\left( E\right) } = \int\limits_{-\infty}^{\infty}
\frac{a_{0}^{\prime }{}^{2}}{2}dx                                \label{III:23}
\end{equation}
is the electric field's energy,
\begin{equation}
E^{\left( G\right) }=\int\limits_{-\infty}^{\infty} \left( f^{\prime }{}^{2}
+s^{\prime }{}^{2}\right)dx                                      \label{III:24}
\end{equation}
is the gradient part of the soliton's energy,
\begin{equation}
E^{\left( T\right) }=\int\limits_{-\infty}^{\infty}
\left( \left( \omega -qa_{0}\right)^{2}s^{2}
+e^{2}a_{0}{}^{2}f^{2}\right) dx                                 \label{III:25}
\end{equation}
is the kinetic part of the soliton's energy, and
\begin{equation}
E^{\left( P\right) } = \int\limits_{-\infty}^{\infty} \left( V\left( f \right)
+ U\left( s \right) \right) dx                                   \label{III:26}
\end{equation}
is the potential part of the soliton's energy.

The energy $E$ of  the  soliton  system  is  the sum of terms (\ref{III:23}) --
(\ref{III:26}).
Using  this  fact  and  virial  relation  (\ref{III:23}),  we  obtain  the  two
representations for the soliton system's energy:
\begin{equation}
E=2\left( E^{\left( T\right) } + E^{\left( G\right) }\right) = 2 \left(
E^{\left( P\right) } + E^{\left( E\right) }\right).              \label{III:27}
\end{equation}
Another representation for  the  soliton  system's  energy  can  be obtained by
integrating the  term  $a_{0}^{\prime }{}^{2}/2$ in Eq.~(\ref{III:15}) by parts
and using  Eqs.~(\ref{III:11}), (\ref{III:14}), and (\ref{III:16}):
\begin{equation}
E = \frac{1}{2}\omega Q_{\chi} + E^{\left( G\right)}
+ E^{\left( P\right)}.                                           \label{III:28}
\end{equation}
Finally, using Eq.~(\ref{III:28}), we obtain the relation  between  the Noether
charge $Q_{\chi}$, the electric field's  energy $E^{\left( E \right)}$, and the
kinetic energy $E^{\left(T\right)}$:
\begin{equation}
\omega Q_{\chi} = 2\left( E^{\left(E\right)} + E^{\left(T\right)} \right).
                                                                 \label{III:29}
\end{equation}

\section{Extreme regimes of the kink-Q-ball system}              \label{sec:IV}

In this section,  we will first study  the properties of the kink-Q-ball system
in the thick-wall regime \cite{kusenko_1997,multam,paccetti}.
In this regime, the parameter $\Delta=\left(m_{\chi}^{2}-\omega^{2}\right)^{1/2}$
tends to zero, so the absolute value of  phase  frequency  tends to $m_{\chi}$.
From   Eqs.~(\ref{III:21})  it  follows  that  in  the  thick-wall  regime, the
functions $s\left(x\right)$  and  $ a_{0}\left(x\right)$  are  spread  over the
one-dimensional space, whereas the  asymptotic  behavior of $f\left( x \right)$
remains unchanged.
In the thick wall regime, functions $s\left(x\right)$ and $ a_{0}\left(x\right)$
uniformly  decrease  as  $\Delta$  and  $\Delta^{2}$,  respectively,  while the
function $f\left( x \right)$  tends  to  non-gauged kink solution (\ref{II:4}).
For this reason, we  perform  the  following scale transformation of the fields
and $x$-coordinate:
\begin{equation}
x = \frac{\bar{x}}{\Delta },\;
s\left( x\right) = \frac{\Delta }{m_{\chi }} \bar{s}\left( \bar{x}\right),\;
a_{0}\left( x\right) = \frac{\Delta^{2}}{m_{\chi }^{2}}
\bar{a}_{0}\left( \bar{x}\right),                                  \label{IV:1}
\end{equation}
while the field $f\left( x \right)$  is  taken  equal  to that of kink solution
(\ref{II:4}).
To research the properties of the kink-Q-ball system  in the thick-wall regime,
we shall use functional (\ref{III:1}) that is related to the energy  functional
through  the  Legendre  transformation: $F\left(\omega\right) = E\left(Q_{\chi}
\right) - \omega Q_{\chi}$.
Using scale  transformation (\ref{IV:1}),  we can determine the leading term of
the dependence  of  the  functional $F\left( \omega\right)$ on $\omega$  in the
thick-wall regime:
\begin{equation}
F\left( \omega \right) = E_{\mathrm{k}} +
\Delta^{3} m_{\chi}^{-2} \bar{F} + O \left(\Delta^{5}\right),      \label{IV:2}
\end{equation}
where $E_{k}= 4\eta ^{3}\sqrt{2\lambda }/3 = 4\eta ^{2}m_{\phi }/3$ is the rest
energy of the non-gauged kink  and  the dimensionless functional $\bar{F}$ does
not depend on $\omega$:
\begin{equation}
\bar{F}=\int\limits_{-\infty }^{\infty }\left[\bar{s}^{\prime
}\left(\bar{x}\right)^{2}+\bar{s}\left(\bar{x}\right)^{2} -
\frac{g_{\chi}}{2m_{\chi}^{2}}\bar{s}\left(\bar{x}\right)^{4}
\right] d\bar{x}.                                                  \label{IV:3}
\end{equation}
In Eq.~(\ref{IV:2}), higher-order terms  in  $\Delta$  may  be neglected in the
thick-wall regime, so we obtain sequentially:
\begin{eqnarray}
Q_{\chi }\left( \omega \right)  &=&-\frac{dF\left( \omega \right) }{d\omega}
=3\bar{F}m_{\chi }^{-2}\omega \left( m_{\chi }^{2}-\omega ^{2}\right)^{
\frac{1}{2}},                                                      \label{IV:4}
\\
E\left( \omega \right)  &=&F\left( \omega \right) -\omega \frac{dF\left(
\omega \right) }{d\omega}  \nonumber
\\
&=&E_{\mathrm{k}} + \bar{F}m_{\chi }^{-2}\left( 2\omega^{2} +
m_{\chi}^{2}\right)\left(m_{\chi}^{2}-\omega^{2}\right)^{\frac{1}{2}},
                                                                   \label{IV:5}
\end{eqnarray}
where known  properties  of  Legendre  transformation are used.
Using Eqs.~(\ref{IV:4}) and (\ref{IV:5}), we obtain the energy of the kink-Q-ball
system as a function of its Noether charge in the thick-wall regime:
\begin{equation}
E = E_{\mathrm{k}}+m_{\chi }Q_{\chi }-\frac{1}{9\times 3!}\frac{m_{\chi}}
{\bar{F}^{2}}Q_{\chi}^{3} + Q\left(Q_{\chi }^{5}\right).           \label{IV:6}
\end{equation}

It has been found numerically  that the  kink-Q-ball  system does not turn into
the  thin-wall  regime  as  the  magnitude  of  the  phase  frequency  tends to
its minimum value.
Such behavior can be qualitatively explained as follows.
Gauss's  law  $a_{0}''\left( x \right) = -j_{0}\left( x \right)$ has an obvious
mechanical analogy.
It describes  a  one-dimensional  motion  of  a  unit  mass  particle along the
coordinate $a_{0}$ in time $x$.
The particle is subjected to  the  time-dependent force $-j_{0}\left(x\right)$.
It starts to move at the time $x = 0$  from the point $a_{0}\left(0\right) > 0$
(we  suppose  that  $\omega > 0$) with  zero  initial  velocity  $a_{0}'\left(0
\right)$.
Since $j_{0}\left(x\right)>0$ as $x < x_{\mathrm{i}}$  ($x_{\mathrm{i}}$ is the
inflection point of $a_{0}\left( x \right)$), the particle's coordinate $a_{0}$
is decreased with increasing of $x$.
At the time  $x_{\mathrm{i}}$,  the   particle  is   at the  point $a_{0}\left(
x_{\mathrm{i}}\right)$ given by Eq.~(\ref{III:21a}) and  possesses the velocity
$a_{0}^{\prime }\left( x_{\mathrm{i}} \right) = - Q_{+}/2$, where $Q_{+} = \int
\nolimits_{-x_{\text{i}}}^{x_{\text{i}}}j_{0}\left(x\right) dx$ is the positive
electric charge of the central area of the soliton system.
The negative electric charge $\int \nolimits_{x_{\text{i}}}^{\infty}j_{0}\left(
x\right) dx$ of  the  side area $x > x_{\mathrm{i}}$ corresponds to the impulse
of force acting in the positive direction.
According to  boundary  conditions  (\ref{III:16a})  and  (\ref{III:16d}), this
impulse of force must bring the particle to rest at the point $a_{0}=0$ as time
$x$ tends to infinity.

As the  magnitude  of  the  phase  frequency $\omega$ approaches to the minimum
value, the  spatial  size  of  the  kink-Q-ball  system  increases, whereas the
positive electric  charge  of  the  central  area  weakly  depends on $\omega$.
As a result, the velocity of the particle at the inflection point $x_{\text{i}}$
remains approximately constant, whereas the  decelerating  force $-j_{0}\left(x
\right)$ decreases because of  the  spreading  of  the  impulse of force $-\int \nolimits_{x_{\text{i}}}^{\infty}j_{0}\left(x\right)dx$ (which is equal to half
the  positive  electric  charge $Q_{+}$ of  the  central  area)  over  the side
area $\left[x_{\text{i}}, \infty \right)$.
Thus the  decelerating   force  $j_{0}\left( x \right)$   decreases  and  so it
cannot stop the particle.
Because of that, the particle reaches the point $a_{0} = 0$ for a finite period
of time, so  boundary  conditions (\ref{III:16a})  and  (\ref{III:16d}) are not
satisfied and the kink-Q-ball system cannot exist.

Next,  we   consider   the   regime   of   small   gauge   coupling  constants.
When the  gauge  coupling  constants  $e$  and  $q$  vanish,  the  gauge  field
$A_{\mu}  = \left( a_{0}\left( x \right),\, 0 \right)$  is  decoupled  from the
kink-Q-ball system,  which  thus  becomes the set of non-gauged kink and Q-ball
that do not interact with each other.
In this connection,  we  want  to ascertain the behavior of the gauge potential
$a_{0}\left( x \right)$ as  $e = \varrho q \rightarrow 0$, where $\varrho$ is a
positive constant.
To do this, we use asymptotic  expressions (\ref{III:19a}) and (\ref{III:21cc})
for $a_{0}\left( x \right)$ that are valid for small and large values of $\left
\vert x \right\vert$, respectively.
We  also  suppose  that   Eqs.~(\ref{III:19a})  and  (\ref{III:21cc})  describe
qualitatively the behavior of $a_{0}$ at intermediate $\left\vert x \right\vert$.
Eqs.~(\ref{III:19a}) and (\ref{III:21cc}) depend on the free parameters $a_{0}$
and $a_{\infty}$, respectively.
These parameters  can   be   determined   by  the  condition  of  continuity of
$a_{0}\left(x\right)$  and  $a_{0}^{\prime}\left(x\right)$ at some intermediate
$x$.
As a result, $a_{0}$ and $a_{\infty}$ become functions of the model's parameters,
including the gauge coupling constants $e$ and $q$.
It can be shown that $a_{0}$ and $a_{\infty}$ tend to the same nonzero limit as
$e = \varrho q \rightarrow 0$.
It follows that at any finite $x$, the gauge  potential $a_{0}\left( x \right)$
tends asymptotically to a constant as $e = \varrho q \rightarrow 0$:
\begin{equation}
\underset{e = \varrho \! q \rightarrow 0}{\lim}
a_{0}\left(x, e, q \right) = \alpha,                               \label{IV:7}
\end{equation}
where the limit value $\alpha$ depends on the model's parameters and $\varrho$.
This fact and linearization  of  Eqs.~(\ref{III:12}) and (\ref{III:13}) lead us
to the asymptotic forms of $f\left( x \right)$ and $s\left( x \right)$ at small
gauge coupling constants:
\begin{subequations} \label{IV:8}
\begin{eqnarray}
f\left( x \right)  & = &f_{\text{k}}\left( x\right) +
e^{2} f_{2}\left( x\right) + O\left( e^{4} \right) ,              \label{IV:8a}
\\
s\left( x \right)  & = &s_{\text{q}}\left( x\right) +
e s_{1}\left( x\right) + O\left( e^{2} \right),                   \label{IV:8b}
\end{eqnarray}
\end{subequations}
where  $f_{\text{k}}\left( x \right)$  and  $s_{\text{q}}\left( x \right)$  are
the non-gauged kink  and  Q-ball solutions, respectively, whereas $f_{2}\left(x
\right)$ and $s_{1}\left( x \right)$  are  some  regular  functions that depend
on the model's parameters (except for $e$ and $q$) and $\varrho$.
Note that  due  to  the  relation  $e  =  \varrho q$, we use only one expansion
parameter $e$.
Substituting  Eqs.~(\ref{IV:7}) and  (\ref{IV:8})  into  Eq.~(\ref{III:14}), we
find the asymptotic behavior of  the  Noether charges $Q_{\phi}$ and $Q_{\chi}$
as $e = \varrho q \rightarrow 0$:
\begin{equation}
Q_{\chi } = -\varrho Q_{\phi } = Q_{0} + e Q_{1} + O\left(e^{2}\right),
                                                                   \label{IV:9}
\end{equation}
where $Q_{0} = 2 \omega \int\nolimits_{-\infty }^{+\infty}s_{\mathrm{q}}^{2}dx$
is the Noehter charge  of  the  non-gauged  Q-ball  and the coefficient $Q_{1}$
depends on  the  model's  parameters  (except  for  $e$ and $q$) and $\varrho$.
Similarly to  Eq.~(\ref{IV:9}),  we  obtain  the  asymptotic   behavior  of the
soliton energy's  components  (\ref{III:23}) -- (\ref{III:26})  and  the  total
soliton energy:
\begin{subequations} \label{IV:10}
\begin{eqnarray}
E^{\left( E\right) } & = & eE_{1}^{\left( E\right) } + O\left(e^{2}\right),
                                                                 \label{IV:10a}
\\
E^{\left( G\right) } & = & E_{0}^{\left( G\right) }+eE_{1}^{\left(
G\right) } + O\left(e^{2}\right),                                \label{IV:10b}
\\
E^{\left( T\right) } & = &E_{0}^{\left( T\right) }+eE_{1}^{\left(
T\right) } + O\left(e^{2}\right),                                \label{IV:10c}
\\
E^{\left( P\right) } & = &E_{0}^{\left( P\right) }+eE_{1}^{\left(
P\right) } + O\left(e^{2}\right),                                \label{IV:10d}
\\
E & = &E_{0}+eE_{1} + O\left(e^{2}\right),                       \label{IV:10e}
\end{eqnarray}
\end{subequations}
where  $E_{0}^{ \left( G\right)} = \int \nolimits_{-\infty }^{ + \infty }\left(
f_{\mathrm{k}}^{\prime 2}+s_{\mathrm{q}}^{\prime 2}\right)dx$, $E_{0}^{\left( T
\right)} = \omega ^{2}\int\nolimits_{-\infty }^{+\infty }s_{\mathrm{q}}^{2}dx$,
and $E_{0}^{\left( P\right) }=\int\nolimits_{-\infty }^{+\infty }\left( V\left(
f_{\mathrm{k}} \right)  +  U\left( s_{\mathrm{q}} \right) \right) dx$  are  the
gradient,  kinetic,  and  potential  parts of  the  non-gauged soliton system's
energy,   respectively,  $E_{0}=E_{0}^{\left( G\right)}+E_{0}^{\left( T\right)}
+E_{0}^{\left(P\right)}$ is the total energy of  the non-gauged soliton system,
and the  coefficients  $E_{1}^{\left( E \right)}$, $E_{1}^{\left( G \right) }$,
$E_{1}^{\left(T\right) }$, and $E_{1}^{\left(P\right)}$ depend  on  the model's
parameters  (except  for  $e$  and $q$) and $\varrho$.
Note that the non-gauged solutions $f_{\mathrm{k}}$ and $s_{\mathrm{q}}$ can be
expressed  in  analytical  form,  as  well  as  the  corresponding energies and
the Noether charges.
The corresponding  expressions  for  the  one-dimensional non-gauged Q-ball are
given in \cite{loginov_prd_99}.
Thus  the  coefficients  $Q_{0}$,  $E_{0}^{\left( G \right)}$,  $E_{0}^{\left(T
\right)}$, $E_{0}^{\left( P  \right)}$,  and  $E_{0}$  can  also  be  expressed
in analytical form.

Next let  us  consider  the  opposite   regime  in  which  both  gauge coupling
constants tend to infinity: $e = \varrho q \rightarrow \infty$.
We  suppose   that    the    behavior    of   the   electromagnetic   potential
$a_{0}\left(x\right)$ in a neighborhood of $x=0$ is regular, so the coefficient
$a_{2}$ given  by  Eq.~(\ref{III:20a}) is  either  finite or tending to zero as
$e = \varrho q \rightarrow \infty$.
But  from   Eqs.~(\ref{III:11})   and   (\ref{III:13})   it   follows  that the
electromagnetic potential is an odd function of $q$:  $a_{0}\left( x, -q\right)
=  -a_{0}\left( x, q\right)$.
Thus, we conclude that  in  the  leading order, $a_{2} \propto q^{-1}$, so from
Eq.~(\ref{III:20a}) it follows that  $a_{0} \sim \omega q^{-1} - a_{-3}q^{-3}$,
where $a_{-3}$ is a positive constant. 
This fact suggests that the electromagnetic potential $a_{0}\left(x\right)$ has
a similar asymptotic expansion in the inverse powers of $e$:
\begin{equation}
a_{0}\left( x\right) = e^{-1}a_{-1}\left( x \right) +
e^{-3}a_{-3}\left(x\right) + O\left(e^{-5}\right),                \label{IV:11}
\end{equation}
where $a_{-1}\left( x \right)$ and $a_{-3}\left( x \right)$  are  some  regular
functions depending  on  the  model's  parameters  (except for $e$ and $q$) and
$\varrho$.
Using  Eqs.~(\ref{III:12}), (\ref{III:13}),  and  (\ref{IV:11}),  we obtain the
general form of asymptotic expansions  for  $f\left( x \right)$  and  $s\left(x
\right)$:
\begin{subequations} \label{IV:12}
\begin{eqnarray}
f\left( x\right)  &  = &f_{0}\left( x\right) + e^{-2}f_{-2}\left( x \right)
+ O\left(e^{-4}\right),                                          \label{IV:12a}
\\
s\left( x\right)  &  = &s_{0}\left( x\right) + e^{-2}s_{-2}\left( x \right)
+ O\left(e^{-4}\right),                                          \label{IV:12b}
\end{eqnarray}
\end{subequations}
where $f_{0}\left( x\right)$,  $f_{-2}\left( x\right)$, $s_{0}\left( x\right)$,
and $s_{-2}\left( x\right)$ are  regular  functions  depending  on  the model's
parameters  (except for $e$ and $q$) and $\varrho$.
Similar to Eqs.~(\ref{IV:7}) and  (\ref{IV:8}),  we  can use Eqs.~(\ref{IV:11})
and (\ref{IV:12}) to obtain the asymptotic  expansions for the soliton energy's
components, the total energy, and the Noether charges:
\begin{subequations} \label{IV:13}
\begin{eqnarray}
E^{\left(E\right)}& = &e^{-2}E_{-2}^{\left( E\right)} +
O\left(e^{-4}\right),                                            \label{IV:13a}
\\
E^{\left( G\right) } & = &\widetilde{E}_{0}^{\left( G\right) } +
e^{-2}E_{-2}^{\left( G\right)} + O\left(e^{-4}\right),           \label{IV:13b}
\\
E^{\left( T\right) } & = &\widetilde{E}_{0}^{\left( T\right) } +
e^{-2}E_{-2}^{\left( T\right)} + O\left(e^{-4}\right),           \label{IV:13c}
\\
E^{\left( P\right) } & = &\widetilde{E}_{0}^{\left( P\right) } +
e^{-2}E_{-2}^{\left( P\right)} + O\left(e^{-4}\right),           \label{IV:13d}
\\
E & = &\widetilde{E}_{0}+e^{-2}E_{-2} + O\left(e^{-4}\right),    \label{IV:13e}
\\
Q_{\chi}&=&-\varrho Q_{\phi} = \widetilde{Q}_{0} + e^{-2}Q_{-2} +
O\left(e^{-4} \right),                                           \label{IV:13f}
\end{eqnarray}
\end{subequations}
where we use the tilde to distinguish corresponding coefficients from  those of
Eqs.~(\ref{IV:10a}) -- (\ref{IV:10d}).
We see that as $e =\varrho q \rightarrow \infty$, the gauge field $a_{0}\left(x
\right)$ tends to zero, so the electric field's energy $E^{\left(E\right)}$ also
vanishes in this regime.
At the same  time, the  products  $e a_{0}\left( x \right)$ and $q a_{0}\left(x
\right)$ tend to nonzero limits $a_{-1}\left( x\right)$ and $\varrho^{-1}a_{-1}
\left( x \right)$,  respectively,  so  the  gauge  field  $a_{0}\left(x\right)$
does not decouple from the kink-Q-ball system.
Due to this, the limit solutions $f_{0}\left(x\right)$ and $s_{0}\left(x\right)$
are different from the corresponding non-gauged solutions $f_{\mathrm{k}}\left(
x\right)$ and $s_{\mathrm{q}}\left(x\right)$, respectively.
From Eqs.~(\ref{IV:13b}) -- (\ref{IV:13f}) it follows that the soliton energy's
components $E^{\left( G\right)}$, $E^{\left( T\right)}$, $E^{\left( P\right)}$,
the total  soliton   energy   $E$,  and  the  Noether  charges  $Q_{\chi}$  and
$Q_{\phi}$ also tend to some finite values as $e=\varrho q \rightarrow \infty$.
It follows that the electric charges  of  the  kink  and  the  Q-ball  increase
indefinitely in  this  regime, despite  the  fact  that  the  electric  field's
energy $E^{\left(E\right)}$ tends to zero.
This is because the electric charges of the  kink and the Q-ball tend to cancel
each other at any spatial point as $e = \varrho q \rightarrow \infty$.
Note that the behavior of the kink-Q-ball system  in  the  extreme regimes $e =
\varrho q \rightarrow 0$ and $e =\varrho q \rightarrow \infty$ was investigated
by numerical methods.
It was found that it is  in  accordance  with Eqs. (\ref{IV:9}), (\ref{IV:10}),
and (\ref{IV:13}).

Finally, we  consider   the   plane-wave  solution of gauge model (\ref{II:1}).
In this case, the gauge field $A^{\mu}$ and the scalar fields $\phi$ and $\chi$
spread  over  the  one-dimensional  space  and fluctuate  around  their  vacuum
values.
Since the  scalar  field  $\phi$  has  nonzero vacuum  value $\left\vert \phi_{
\text{vac}}\right\vert = \eta$, the  classical  vacuum of model (\ref{II:1}) is
not invariant under  local  gauge  transformations  (\ref{II:4a}), so the local
gauge symmetry is spontaneously broken.
For this  reason,  the  research  of  the  plane-wave solution is convenient to
perform in the unitary gauge $\text{Im}\left(\phi\left(x,t\right) \right) = 0$.
In this gauge, the Higgs mechanism is  realized  explicitly, so we can read off
the particle  composition  of  model  (\ref{II:1}).
In  the  neighborhood   of   the   gauge  vacuum  $\phi_{\text{vac}}  =  \eta$,
$\chi_{\text{vac}} = 0$,  we  have  the  complex  scalar  field $\chi$ with the
mass $m_{\chi}$, the real scalar Higgs field $\phi_{H}$ with the mass $m_{\phi}
=\sqrt{2\lambda}\eta$, and the massive  gauge field   $A^{\mu}$  with the  mass
$m_{A} = \sqrt{2} e \eta$.

We want to find the spatially uniform solution of field equations (\ref{II:7a})
-- (\ref{II:7c})  possessing  the  Noether  charges  $Q_{\phi}$  and $Q_{\chi}$
(recall that $e Q_{\phi}+q Q_{\chi}=0$ for any finite energy field configuration)
and to determine its energy.
For this, we use field  equations (\ref{II:7a}) -- (\ref{II:7c}) in the unitary
gauge.
We suppose that with unlimited spreading, the  amplitudes of the complex scalar
field $\chi$  and  the  real  scalar  Higgs  field $\phi_{H}$  tend to zero, so
we can neglect higher-order terms in the Lagrangian (\ref{II:1}).
On spatially uniform fields,  the  field equations for $A^{\mu}$ and $\phi_{H}$
become algebraic  ones,  whereas  the  field equation for $\chi$ determines the
time dependence of $\chi$.
The results  obtained  are  presented  as  series  in  inverse  powers  of  the
plane-wave solution's spatial size $L$:
\begin{equation}
A_{\text{pw}}^{\mu } = \left( \frac{q\lambda }{e^{2}m_{\phi }^{2}}
\frac{Q_{\chi }}{L}+O\left( \frac{1}{L^{3}}\right)\!,\,0\right)\!,
                                                                 \label{IV:14a}
\end{equation}
\begin{equation}
\phi_{H\text{pw}} = \frac{q^{2}}{e^{2}}\frac{\lambda ^{3/2}}
{\sqrt{2}m_{\phi }^{5}}\frac{Q_{\chi }^{2}}{L^{2}}+
O\left( \frac{1}{L^{4}}\right)\!,
                                                                 \label{IV:14b}
\end{equation}
\begin{eqnarray}
\chi_{\text{pw}} &=&\sqrt{\frac{Q_{\chi }}{2m_{\chi }L}}
\left(1+O\left( \frac{1}{L^{4}}\right) \right)
                                                                 \label{IV:14c}
\\
&&\times \exp \left[ -i\left( m_{\chi }+\frac{q^{2}}{e^{2}}\frac{\lambda }
{m_{\phi }^{2}}\frac{Q_{\chi }}{L}+O\left( \frac{1}{L^{3}}\right) \right) t
\right]. \nonumber
\end{eqnarray}
We see that as $L \rightarrow \infty$, the amplitudes  of  the  fields $A^{0}$,
$\phi_{H}$, and $\chi$  tend  to  zero,  whereas the phase frequency of  $\chi$
tends to $m_{\chi}$. 
Note that the fields $A^{0}$, $\phi_{H}$, and $\chi$ of the plane-wave solution
tend to zero as $L^{-1}$, $L^{-2}$, and $L^{-1/2}$, respectively,  so the Higgs
field $\phi_{H}$ tends to zero much more quickly  than the complex scalar field
$\chi$.
Substituting  Eqs.~(\ref{IV:14a}) -- (\ref{IV:14c}) into  Eqs.~(\ref{II:6}), we
obtain the  Noether  charge  densities $j_{\phi}^{0}$ and $j_{\chi}^{0}$ of the
plane-wave solution:
\begin{eqnarray}
j_{\phi}^{0}&=&-\frac{q}{e}\frac{Q_{\chi}}{L}+O\left(\frac{1}{L^{5}}\right),
                                                                 \label{IV:15a}
\\
j_{\chi}^{0}&=&\frac{Q_{\chi}}{L}+O\left(\frac{1}{L^{5}}\right). \label{IV:15b}
\end{eqnarray}
From Eqs.~(\ref{IV:15a}) and (\ref{IV:15b}) it follows that the electric charge
of the plane-wave solution vanishes:
\begin{equation}
Q=\underset{L\rightarrow \infty }{\lim }L\left(e j_{\phi }^{0} +
q j_{\chi}^{0}\right) = 0,                                        \label{IV:16}
\end{equation}
as it should be.
Next, we  calculate  the  energies  of  the $\phi$ and $\chi$ components of the
plane-wave solution:
\begin{equation}
E_{\phi }=\frac{q^{2}}{e^{2}}\frac{\lambda }{2m_{\phi }^{2}}
\frac{Q_{\chi }^{2}}{L}+O\left( \frac{1}{L^{3}}\right),           \label{IV:17}
\end{equation}
and
\begin{equation}
E_{\chi }=m_{\chi }Q_{\chi }+O\left(L^{-4}\right),                \label{IV:18}
\end{equation}
so the  total  energy  of  the  plane-wave  solution  turns  out to be equal to
\begin{equation}
E_{\mathrm{pw}}=\underset{L\rightarrow \infty }{\lim}\left( E_{\phi} +
                E_{\chi} \right) = m_{\chi}Q_{\chi}.              \label{IV:19}
\end{equation}

Let us discuss the results obtained.
First of  all, from  Eqs.~(\ref{IV:17}) -- (\ref{IV:19})  it  follows  that the
$\phi$ component does not contribute to the energy of the  plane-wave  solution
as $L \rightarrow \infty$.
At the same time, from Eqs.~(\ref{IV:15a}) and (\ref{IV:16}) it follows that the
electric charge of the $\phi$ component does not vanish and is opposite to that
of the  $\chi$  component,  so  the  total  electric  charge  of the plane-wave
solution vanishes.
Thus, the $\phi$ component contributes to the electric charge of the plane-wave
solution, but does not contribute to its energy.
This can be explained as follows.
In  the  unitary  gauge  $\text{Im}\left(  \phi \right) = 0$,  the  real  Higgs
field $\phi_{H}$  fluctuates  around  the  real  vacuum  average $\eta$, so the
initial scalar field $\phi$ is written as $\phi = \eta + \phi_{H}$.
Further, from  Eqs.~(\ref{II:6a})  and  (\ref{II:10})  we  obtain  the electric
charge and energy densities of the $\phi$ component of the plane-wave solution:
\begin{equation}
j_{\phi}^{0} = -2 e A^{0}\left(\eta +\phi_{H}\right)^{2} \sim -
2 e \eta^{2} A^{0}                                                \label{IV:20}
\end{equation}
and
\begin{eqnarray}
\mathcal{E}_{\phi} & = &\left( D_{0}\phi \right) ^{\ast }D_{0}\phi
+ m_{\phi}^{2}\phi_{H}^{2} \nonumber \\
&=& e^{2}A_{0}^{2}\left( \eta +\phi_{H}\right)^{2}+m_{\phi }^{2}\phi
_{H}^{2}\sim e^{2}\eta ^{2}A_{0}^{2},                             \label{IV:21}
\end{eqnarray}
where Eqs.~(\ref{IV:14a}) and (\ref{IV:14b}) have been used.
We see that in  the  leading  order  in  $L^{-1}$, the  Higgs  field $\phi_{H}$
contributes  neither  to  $j_{\phi}^{0}$  nor  to $\mathcal{E}_{\phi}$, whereas
the vacuum  average  of  the  scalar  field  $\phi$  contributes in both cases.
We also see that as $L \rightarrow \infty$, the behavior of  $j_{\phi}^{0}$ and
$\mathcal{E}_{\phi}$  is  determined  only  by  the  electromagnetic  potential
$A^{0}$.
At the same time, from  Eq.~(\ref{IV:14a})  it follows  that for the plane-wave
solution, $A^{0} \sim L^{-1}$, so $j_{\phi}^{0} \sim L^{-1}$ and $\mathcal{E}_{
\phi} \sim L^{-2}$ in the leading order in $L^{-1}$.
It follows  that  $Q_{\phi} = \int\nolimits_{-\infty}^{+\infty}j_{\phi}^{0}dx =
O\left(1\right)$ and $E_{\phi} = \int\nolimits_{-\infty}^{+\infty}\mathcal{E}_{
\phi }dx = O\left(L^{-1}\right)$,  so  the $\phi$ component does not contribute
to the plane-wave solution's energy as $L \rightarrow \infty$.

\section{Numerical results}                                       \label{sec:V}

To study the kink-Q-ball system, we must solve system of differential equations
(\ref{III:11}) -- (\ref{III:13}) satisfying boundary conditions (\ref{III:16}).
This  first   boundary   value   problem   can   be  solved  only  numerically.
To solve the boundary  value  problem,  we use the method of finite differences
and subsequent  Newtonian  iterations   realized   in  the {\sc{Maple}} package
\cite{maple}.
To check  the  correctness of a numerical solution, we use Eqs.~(\ref{III:10}),
(\ref{III:16e}), and (\ref{III:22}).

To solve the boundary value  problem,  we  need to  know  the eight dimensional
parameters:  $\omega$, $e$, $q$, $m_{\phi} = \sqrt{2 \lambda} \eta$, $\lambda$,
$m_{\chi}$, $g_{\chi}$, and $h_{\chi}$.
Without loss of generality  the  mass  $m_{\chi}$  can  be chosen as the energy
unit, so the dimensionless  functions $a_{0}\left(x\right)$, $f\left(x\right)$,
and  $s\left(x\right)$  depend  only  on  the  seven  dimensionless  parameters
$\tilde{\omega}  =  \omega/m_{\chi}$,  $\tilde{e}  = e/m_{\chi}$,  $\tilde{q} =
q/m_{ \chi }$,  $\tilde{m}_{ \phi}  =  m_{ \phi}/m_{\chi}$,  $\tilde{\lambda} =
\lambda/m_{\chi}^{2}$, $\tilde{g}_{\chi} = g_{\chi}/m_{\chi}^{2}$, and $\tilde{
h}_{\chi} = h_{\chi}/m_{\chi}^{2}$.
In the  present  paper,  we  consider  the  kink-Q-ball  system  for  which the
dimensionless non-gauged parameters $\tilde{m}_{\phi}=\sqrt{2}$, $\tilde{\lambda}
= 1$, $\tilde{g}_{\chi} = 2.3$,  and  $\tilde{h}_{\chi}  =  1$  are of the same
order of magnitude.
The dimensionless gauge  coupling  constants  $\tilde{e}$  and $\tilde{q}$ were
taken to be equal to  each  other  and  could  take  the  values $0.05$, $0.1$,
$0.2$, $0.3$, $0.4$, and $0.5$.

\begin{figure}[tbp]
\includegraphics[width=0.5\textwidth]{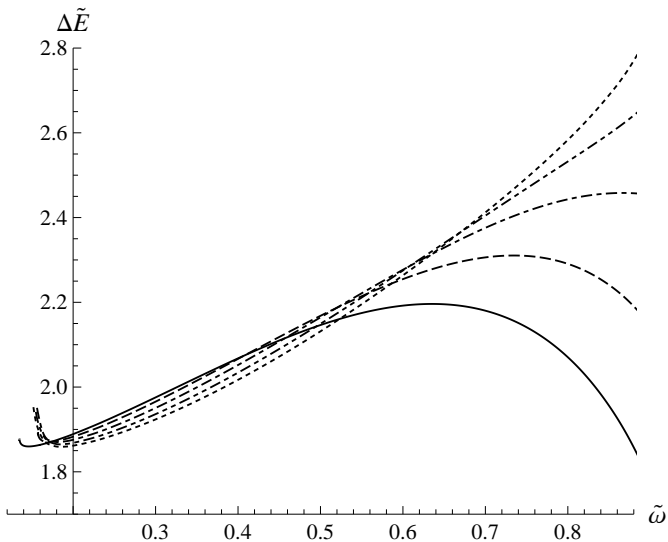}
\caption{The dependences of the dimensionless energy difference $\Delta\tilde{E
}$ on the  dimensionless  phase frequency $\tilde{\omega}$.  The solid, dashed,
dash-dotted,  dash-dot-dotted,  and  dotted  curves  correspond  to  the  gauge
coupling  constants  $\tilde{e}  =  \tilde{q}  =  0.1$, 0.2, 0.3, 0.4, and 0.5,
respectively}
\label{fig:1}
\end{figure}

\begin{figure}[tbp]
\includegraphics[width=0.5\textwidth]{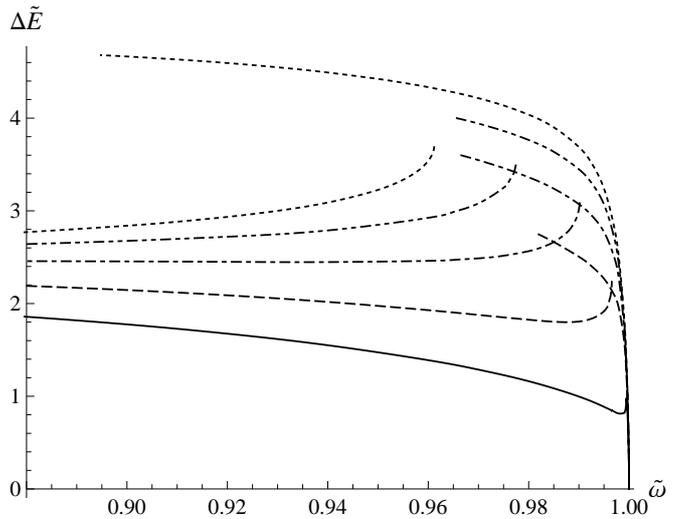}
\caption{The dependences of the dimensionless energy difference $\Delta\tilde{E
}$ on the  dimensionless  phase  frequency  $\tilde{\omega}$.  The notations of
curves are the same as in Fig.~\ref{fig:1}}
\label{fig:2}
\end{figure}

Let us  denote  by  $\Delta E$  the  difference  between  the  energies  of the
kink-Q-boll  system  and  the  non-gauged  kink: $\Delta E = E - E_{\text{k}}$,
where $E_{\mathrm{k}}=4\eta ^{3}\sqrt{2\lambda }/3$.
Figures~\ref{fig:1} and \ref{fig:2} present the dependence of the dimensionless
energy difference $\Delta\tilde{E}=\Delta E/m_{\chi}$ on the dimensionless phase
frequency $\tilde{\omega}$.
The curves in these figures correspond to gauge coupling constants $\tilde{e} =
\tilde{q}$ taking the values from the set $0.1$, $0.2$, $0.3$, $0.4$, and $0.5$.
Figure~\ref{fig:1} presents  the curves in the range from the minimum values of
$\tilde{\omega}$, which we  managed to reach by numerical methods, to the value
$\tilde{\omega} = 0.88$.
Figure~\ref{fig:2} presents the same curves in the range from $\tilde{\omega} =
0.88$ to the maximum possible value $\tilde{\omega} = 1$.
We use the two figures for  a  better representation of the dependences $\Delta
\tilde{E}\left(\tilde{\omega}\right)$.
For the same values of gauge coupling constants, the dependences $\Delta\tilde{
E}\left( \tilde{\omega} \right)$ and $Q_{\chi}\left(\tilde{\omega} \right)$ are
qualitatively similar, so the dependences $Q_{\chi}\left(\tilde{\omega}\right)$
are not given in the present paper.

Let us  discuss  the  main  features  of  the  curves  in Figs.~\ref{fig:1} and
\ref{fig:2}.
First of all, we note that the energy of the kink-Q-ball system  does  not tend
to infinity as $\tilde{\omega}$ tends to its minimum values (that depend on the
gauge coupling constants).
Indeed, it was found numerically that  the  dependences $\tilde{E}\left(\tilde{
\omega} \right)$  and  $Q_{\chi}\left( \tilde{\omega} \right)$ have a branching
point at $\tilde{\omega}_{\text{min}}$:
\begin{eqnarray}
\tilde{E} &\sim &A-B\tilde{\omega }_{\min }\left( \tilde{\omega}
-\tilde{\omega}_{\min }\right)^{1/2},                              \label{V:1a}
\\
Q_{\chi } &\sim &C-B\left(\tilde{\omega} - \tilde{\omega}_{\min}
\right)^{1/2},                                                     \label{V:1b}
\end{eqnarray}
where $A$, $B$, and $C$ are positive constants.
We were unable to find any solutions of the boundary value problem for $\tilde{
\omega} < \tilde{\omega}_{\min}$,  so  we  conclude that the kink-Q-ball system
does not  turn  into  the  thin-wall  regime  in  which  both  $\tilde{E}$  and
$Q_{\chi}$ must tend to infinity.

\begin{figure}[b]
\includegraphics[width=0.5\textwidth]{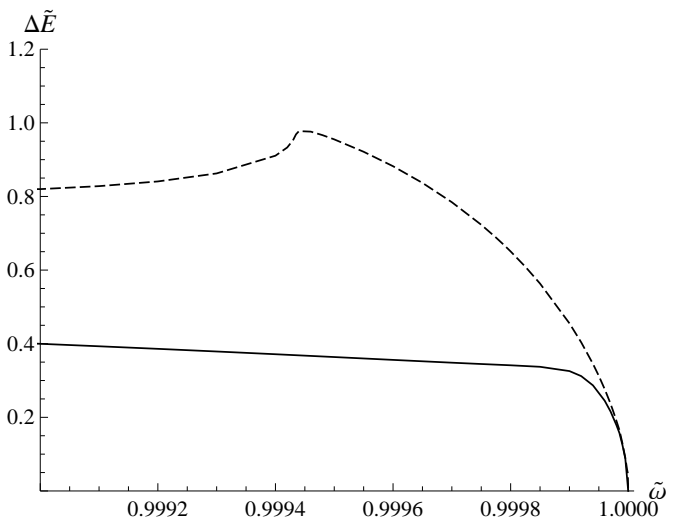}
\caption{The dependences of the dimensionless energy difference $\Delta\tilde{E
}$ on the dimensionless  phase  frequency  $\tilde{\omega}$ in the neighborhood
of $\tilde{\omega} = 1$.  The solid and dashed curves correspond  to  the gauge
coupling  constants $\tilde{e}  =  \tilde{q}  =  0.05$ and $0.1$, respectively}
\label{fig:3}
\end{figure}

The behavior of the curves in the neighborhood of $\tilde{\omega} = 1$  is also
rather unusual.
We see that for the gauge coupling constants $\tilde{e}=\tilde{q}$ from the set
$0.2$, $0.3$, $0.4$, and $0.5$, the  dependence  $\tilde{E}\left(\tilde{\omega}
\right)$ consists of two separate curves.
The left curve starts from the  minimal phase frequency $\tilde{\omega}_{\min}$
(where it has branching point (\ref{V:1a})) and  continues  until  the  maximal
phase frequency $\tilde{\omega}_{\text{r}}$.
The behavior of the left curve in the neighborhood of $\tilde{\omega}_{\text{r}
}$ is similar to that in the neighborhood of $\tilde{\omega}_{\min}$:
\begin{eqnarray}
\tilde{E} &\sim &D - F \tilde{\omega}_{\text{r}}\left(\tilde{\omega}_{\text{r}}
-\tilde{\omega} \right)^{1/2},                                     \label{V:2a}
\\
Q_{\chi } &\sim &G - F \left(\tilde{\omega}_{\text{r}} - \tilde{\omega}
\right)^{1/2},                                                     \label{V:2b}
\end{eqnarray}
where $D$, $F$, and $G$ are positive constants.
The right  curve starts from  some  phase  frequency $\tilde{\omega}_{\text{l}}
< \tilde{\omega}_{\text{r}}$ and  continues  up  to  the maximum possible value
$\tilde{\omega}_{\text{tk}} = 1$.
The curve has no singularity in the neighborhood of $\tilde{\omega}_{\text{l}}$.
According to Seq.~\ref{sec:V}, the kink-Q-ball system  goes into the thick-wall
regime as $\tilde{\omega} \rightarrow 1$.
Indeed, it was found numerically that in the  neighborhood of $\tilde{\omega}_{
\text{tk}} = 1$, $\Delta\tilde{E} \sim Q_{\chi} \sim  H \left( \tilde{\omega}_{
\text{tk}}  - \tilde{\omega}\right)^{1/2}$ in accordance with Eqs.~(\ref{IV:4})
and (\ref{IV:5}).

Figure~\ref{fig:3}  shows the  dependences $\Delta\tilde{E}\left(\tilde{\omega}
\right)$ for $\tilde{e} = \tilde{q} = 0.05$  and  $\tilde{e} = \tilde{q} = 0.1$
in the neighborhood of $\tilde{\omega}_{\text{tk}} = 1$.
We see that with decreasing gauge coupling constants, the left and right curves
are merged into one, so  the  dependence  $\Delta\tilde{E}\left( \tilde{\omega}
\right)$ becomes single-valued.
In accordance  with  Seq.~\ref{sec:IV},  both  curves  go  into  the thick-wall
regime as $\tilde{\omega} \rightarrow 1$.

\begin{figure}[t]
\includegraphics[width=0.5\textwidth]{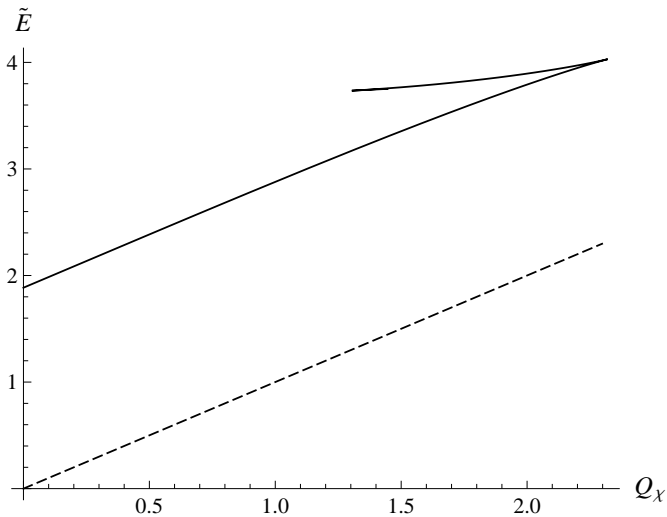}
\caption{The dependence  of  the  dimensionless   energy   $\tilde{E}$  of  the
kink-Q-ball system  with $\tilde{e} = \tilde{q} = 0.05$ on  the  Noether charge
$Q_{\chi}$ (solid curve). The dashed line $\tilde{E} = Q_{\chi}$ corresponds to
the plane-wave solution}
\label{fig:4}
\end{figure}

\begin{figure}[t]
\includegraphics[width=0.5\textwidth]{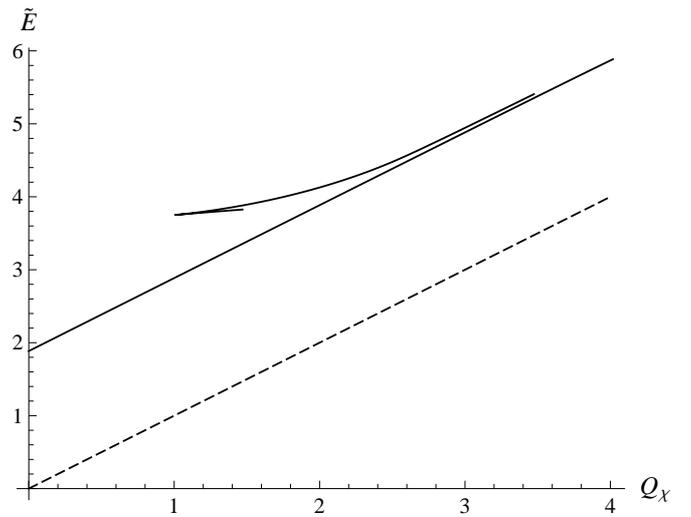}
\caption{The dependence  of  the  dimensionless   energy   $\tilde{E}$  of  the
kink-Q-ball  system  with $\tilde{e} = \tilde{q} = 0.4$ on  the  Noether charge
$Q_{\chi}$ (solid curve). The dashed line $\tilde{E} = Q_{\chi}$ corresponds to
the plane-wave solution}
\label{fig:5}
\end{figure}

Knowing the  dependences  $\tilde{E}\left( \tilde{\omega} \right)$ and $Q_{\chi
}\left(\tilde{\omega}\right)$, we can obtain the dependence $\tilde{E}\left(Q_{
\chi}\right)$.
Figures~\ref{fig:4}  and  \ref{fig:5} show  the  dependence $\tilde{E}\left(Q_{
\chi}\right)$ for the gauge  coupling  constants $\tilde{e} = \tilde{q} = 0.05$
and $\tilde{e} = \tilde{q} = 0.4$, respectively.
The straight  lines  $\tilde{E} = Q_{\chi}$  in these figures correspond to the
plane-wave solution.
We see that for $\tilde{e} = \tilde{q} = 0.05$, the dependence $\tilde{E}\left(
Q_{\chi}\right)$ is a single connected curve, whereas for $\tilde{e} =\tilde{q}
= 0.4$, it consists of two separate curves.
Of course,  the  number  of  curves  in  Figs.~\ref{fig:4}  and  \ref{fig:5} is
determined by the  number  of  the  corresponding  curves in Figs.~\ref{fig:2}.
The curves in Figs.~\ref{fig:4}  and \ref{fig:5} possess cusps, whose number is
determined  by  the   number   of  extremes  of  the  corresponding  curves  in
Figs.~\ref{fig:1}  and  \ref{fig:2}.
The second  derivative  $d^{2}\tilde{E}/dQ_{\chi }^{2}$  changes  the sign when
passing through the cusps or  discontinuities,  so  convex and concave sections
of the curves change each other.
Note that in Figs.~\ref{fig:4} and \ref{fig:5},  the  energy of the kink-Q-ball
system turns out to be more than the energy of the plane-wave solution with the
same value of $Q_{\chi}$.
This also turns out to be  true  for  all other cases considered in the present
paper.
It follows that the  kink-Q-ball  system  may transit into the plane-wave field
configuration through quantum tunneling.

\begin{figure}[t]
\includegraphics[width=0.5\textwidth]{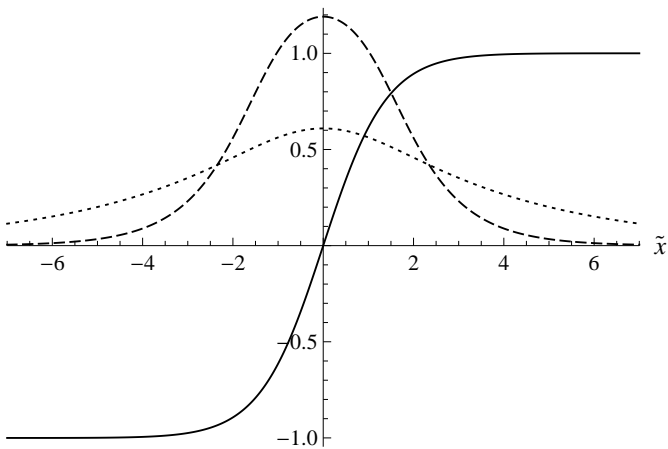}
\caption{The  numerical  kink-Q-ball  solution  corresponding  to  $\tilde{e} =
\tilde{q} = 0.2$  and  $\tilde{\omega} = 0.3$.    The solid, dashed, and dotted
curves correspond to  $f\left(\tilde{x}\right)$, $s\left(\tilde{x}\right)$, and
$a_{0}\left(\tilde{x}\right)$, respectively}
\label{fig:6}
\end{figure}

\begin{figure}[t]
\includegraphics[width=0.5\textwidth]{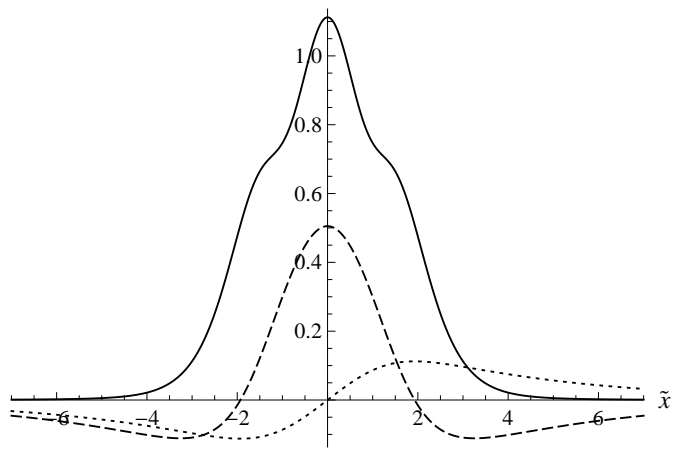}
\caption{The dimensionless versions of the  energy density $\tilde{\mathcal{E}}
= m_{\chi}^{-2}\mathcal{E}$, (solid  curve), the scaled electric charge density
$\tilde{e}^{-1}\tilde{j_{0}}=\tilde{e}^{-1} m_{\chi}^{-2}j_{0}$ (dashed curve),
and the electric  field  strength $\tilde{E}_{x} = m_{\chi}^{-1} E_{x}$ (dotted
curve), corresponding to the kink-Q-ball solution in Fig.~6}
\label{fig:7}
\end{figure}

Figure~\ref{fig:6} shows the kink-Q-ball solution  with the dimensionless phase
frequency $\tilde{\omega} = 0.3$.
The energy density, the electric charge density, and the electric field strength
corresponding to this  kink-Q-ball  solution are presented in Fig.~\ref{fig:7}.
We see that in accordance with  Seq.~\ref{sec:III}, $a_{0}\left( x \right)$ and
$s\left( x\right)$ are even functions  of $x$, whereas $f\left( x\right)$ is an
odd function of $x$.
We also see that $f\left( x\right)$ and $s\left( x \right)$ reach neighborhoods
of their boundary  values (\ref{III:16})  faster  than $a_{0}\left( x \right)$.
From Fig.~\ref{fig:7}  it  follows  that  the  kink-Q-ball system possesses the
symmetric energy and electric charge densities.
It also possesses the nonzero electric  field  strength that is an odd function
of the space coordinate.
The distribution of the electric charge density is a central symmetric peak with
a positive  $j_{0}$  surrounded  by  two  areas with a negative $j_{0}$, so the
total electric charge of the kink-Q-ball system vanishes. 
The central positive peak  is  due  to  the  contribution  of the field $\chi$,
whereas the two side negative areas are due to  the  contribution  of the field
$\phi$.
Note that the energy density $\mathcal{E}$ reaches a close neighborhood of zero
faster than the electric charge density $j_{0}$ and the electric field strength
$E_{x}$.
The reason is the similar behavior of  the electromagnetic potential $a_{0}$ in
Fig.~\ref{fig:6}.
Indeed, from Eq.~(\ref{III:14}) it follows that the  electric charge density of
the field $\phi$ is $-2a_{0}e^{2}\!f^{2}$.
We see that the electromagnetic potential $a_{0}$ can induce a nonzero electric
charge density even if the scalar field $\phi$  reaches a close neighborhood of
the vacuum value $\left\vert \eta \right\vert$.
As a result, a substantial  part of the  electric  charge of the complex scalar
field $\phi$ comes from the parts of the two  side areas where $\left\vert \phi
\right\vert  \approx  \left\vert  \eta  \right\vert$,  $\chi  \approx  0$,  and
$\mathcal{E} \approx 0$.

Let us now discuss the issue of stability of the kink-Q-ball system.
As has already been pointed  out,  the  energy of the kink-Q-ball system turned
out to  be  more  than  the  energy  of  the  plane-wave solution with the same
$Q_{\chi}$ for all cases considered in the present paper.
It follows that the  kink-Q-ball  system  is  unstable  against  transit into a
plane-wave configuration through quantum tunneling.
It remains to consider the stability of the  kink-Q-ball system with respect to
classic fluctuations of the fields $\phi$, $\chi$, and $A_{\mu}$ in a functional
neighborhood of the kink-Q-ball solution.

It is known that the gauge model described by the  first line of the Lagrangian
(\ref{II:1})  possesses  the  kink  solution  \cite{bochkarev_mpl_2,grigoriev}.
In  the  adopted  gauge   $A_{x}  =  0$,   this   kink  solution  is  given  by
Eq.~(\ref{II:4}).
The gauged kink  has  zero  electric  charge,  so  it  possesses finite energy.
However,  unlike   the   kink   of   a   self-interacting   real  scalar  field
\cite{dashen,polyakov}, the  gauged  kink  is  not a topologically stable field
configuration.
Due to the topological structure of the  vacuum of the Abelian Higgs model, the
gauged kink is a sphaleron \cite{bochkarev_mpl_2,grigoriev}.
The existence  of  the  sphaleron  is  due to the paths in the functional space
that  connect  topologically  distinct  vacua   of   the  Abelian  Higgs  model
\cite{Manton}.
The sphaleron lies  between  two  topologically  distinct neighboring vacua and
has {\it{exactly one}} unstable mode.

The gauged kink is a static solution  modulo gauge transformations.
However, in the case of the kink-Q-ball solution, we have a different situation.
It can easily be shown that the kink-Q-ball solution will depend on time in any
gauge, so it is not a static solution.
It follows  that  the  point  in  the  functional  space  corresponding  to the
kink-Q-ball solution will vary with time in any gauge.
This fact does  not  allow  the  kink-Q-ball solution to be a sphaleron, so the
question about  the  unstable  modes  of  the  kink-Q-ball  solution  should be
investigated separately.

To investigate the classic stability of the kink-Q-ball system, it is necessary
to study the spectrum of the operator  of second variational derivatives in the
functional neighborhood of the kink-Q-ball solution.
Wherein, the  model's  fields  must  fluctuate  so  that  the  Noether  charges
$Q_{\phi}$ and  $Q_{\chi}$  remain  fixed  and  the  perturbed  electromagnetic
potential $A_{0} + \delta A_{0}$ continues to satisfy Gauss's law.
All these factors make it difficult to  study the spectrum even using numerical
methods.
However, these  difficulties  can  be  avoided  if  we  numerically solve field
equations (\ref{II:7a}) -- (\ref{II:7c}) in  the temporal gauge $A_{0} = 0$ and
with a perturbed initial  field  configuration in the close neighborhood of the
kink-Q-ball solution.
Indeed, Gauss's law can be easily implemented in the temporal gauge at $t = 0$,
whereupon Gauss's  law  will be automatically satisfied for $t > 0$.
The perturbed initial  field  configuration   must have the same $Q_{\phi}$ and
$Q_{\chi}$ as the kink-Q-ball system; then the field  equations  guarantee that
the field  configuration  will  also have the same $Q_{\phi}$ and $Q_{\chi}$ at
later times.
Having perturbed and unperturbed kink-Q-ball solutions of  the field equations,
we can observe how field fluctuations behave as time increases.
If any  fluctuation  of  fields  oscillates  in  a  close  neighborhood  of the
kink-Q-ball solution then the solution is classically stable.
If there exists at least one fluctuation of fields that increases exponentially
with time then the kink-Q-ball solution is classically unstable.

In the present paper, we research the  stability  of the kink-Q-ball system for
$\tilde{e}  =  \tilde{q}  =  0.05$,  $0.1$,  $0.2$,  $0.3$,  $0.4$,  and $0.5$.
For each value  of  the  gauge coupling constants, we took the step of changing
of $\tilde{\omega}$ equal to $0.1$.
To solve  field  equations (\ref{II:7a}) -- (\ref{II:7c}),  we  use  the solver
of  partial  differential  equations  realized   in  the  {\sc{Maple}}  package
\cite{maple}.
We have found that there are at least two unstable modes for the all considered
gauge coupling constants and phase frequencies.
The first  unstable  mode  corresponds  to an initial symmetric perturbation of
$\text{Im}\phi$, whereas the second one corresponds to an initial antisymmetric
perturbation of $\text{Im} \chi$.
It follows that the kink-Q-ball system is not a sphaleron since it has at least
two unstable modes,  whereas  a  sphaleron must have exactly one unstable mode.

\section{Conclusion}                                             \label{sec:VI}

In the present paper, we consider one-dimensional model (\ref{II:1}) consisting
of two self-interacting  complex  scalar fields interacting through an  Abelian
gauge field.
It was shown  that  the  model  possesses  the soliton solution consisting of a
gauged kink and a gauged Q-ball.
Since the finiteness of the energy of the  one-dimensional soliton system leads
to its electric  neutrality,  the  gauged  kink  and  the  gauged  Q-ball  have
opposite electric charges.
Due to the neutrality of the Abelian gauge field, the opposite electric charges
of the kink and Q-ball components are conserved separately.
Despite the  neutrality  of  the  kink-Q-ball  system,  it  possesses a nonzero
electric field.

The kink-Q-ball system  has  rather  unusual  dependences of the energy and the
Noether charge on the phase frequency.
Indeed, it was found that the energy  and the Noether charge of the kink-Q-ball
system do not tend to  infinity  as  the  phase  frequency tends to its minimum
value, but instead have the branch point.
It follows  that  there  is  no  thin-wall  regime  for the kink-Q-ball system.
We also  found  that  when  the  magnitude  of  the  phase  frequency is in the
neighborhood  of  $m_{\chi}$, the  dependences  of  the  energy and the Noether
charge on the  phase  frequency  consist of two separate branches provided that
the model's gauge coupling constants are large enough.
In all cases, however, the  kink-Q-ball  system goes into the thick-wall regime
as the magnitude  of  the  phase  frequency tends to $m_{\chi}$.

In addition to the kink-Q-ball solution,  the model also possesses a plane-wave
solution.
For all  sets  of  the  model  parameters  considered in the present paper, the
energy of the  kink-Q-ball solution turns out to be more than the energy of the
plane-wave solution with the same value of the Noether charge.
Due to the topological structure of the model's vacuum, the kink-Q-ball solution
is  not  topologically   stable,  so   it   can  transit  into  the  plane-wave
configuration through quantum tunneling.

It is known that the  Abelian  Higgs  model  possesses the gauge kink solution.
The gauge kink  is   electrically   neutral   and   has   one   unstable  mode.
From the viewpoint of  topology,  the  gauge  kink  is  a  static (modulo gauge
transformations)  field  configuration  lying  between  the  two  topologically
distinct adjacent vacua.
Unlike this, the  kink-Q-ball  solution  depends on time in any gauge, so it is
not a static field configuration.
Hence, the kink-Q-ball solution cannot  be  a  sphaleron,  and  so the question
of its classic stability requires separate consideration.
We research  the  classic  stability  of  the  kink-Q-ball  system  by means of
numerical solution of the  field  equations  with  initial field configurations
perturbed in the close neighborhood of the kink-Q-ball solution.
It was found that in all  considered  cases,  the  kink-Q-ball  solution has at
least two unstable  modes,  so  it  is even more unstable than the gauged kink.

\begin{acknowledgements}

This work was supported by the Russian Science Foundation, grant No 19-11-00005.

\end{acknowledgements}

\appendix

\section{Quasi-classical limit of the soliton system}              \label{app1}

In this Appendix,  we  consider  the  quasi-classical limit of  the kink-Q-ball
system.
In this limit, quantum field fluctuations are localized in a close neighborhood
of the classical solution, so the   soliton   system   behaves like a classical
object.
Following \cite{lee}, we  form  from  the  coupling  constants  $g_{\chi}$  and
$h_{\chi}$ the  two  dimensionless  combinations: $\epsilon = \left[ 16h_{\chi}
m_{\chi }^{2}/\left( 3 g_{\chi }^{2}\right) - 1 \right]^{1/2}$ and $g = 2\left[
h_{\chi }/\left( 3g_{\chi }\right) \right]^{1/2}$.
Let us remember that the condition $3g_{\chi }^{2}<16h_{\chi }m_{\chi }^{2}$ is
assumed to be fulfilled, so the parameter $\epsilon$ is real.
The coupling constants $g_{\chi}$ and $h_{\chi}$  are expressed in terms of $g$
and $\epsilon$ as follows:
\begin{equation}
g_{\chi }=g^{2}\bar{g}_{\chi },\;h_{\chi }=g^{4}\bar{h}_{\chi },    \label{A:1}
\end{equation}
where the rescaled coupling constants $\bar{g}_{\chi}$ and $\bar{h}_{\chi}$ are
\begin{eqnarray}
\bar{g}_{\chi } &=&\frac{4m_{\chi }^{2}}{1+\epsilon^{2}}=\frac{3}{4}
\frac{g_{\chi }^{2}}{h_{\chi }},                                   \label{A:2a}
\\
\bar{h}_{\chi } &=&\frac{3m_{\chi }^{2}}{1+\epsilon ^{2}}=\frac{9}{16}
\frac{g_{\chi }^{2}}{h_{\chi }},                                   \label{A:2b}
\end{eqnarray}
so $\bar{h}_{\chi} = 3\bar{g}_{\chi}/4$.
Next, we use the dimensionless  parameter $g$ to rescale the model's fields and
remaining coupling constants as follows: $\phi=g^{-1}\bar{\phi}$, $\chi =g^{-1}
\bar{\chi}$,   $\eta  =  g^{-1}\bar{\eta}$,  $A^{\mu}  =  g^{-1}\bar{A}^{\mu}$,
$\lambda = g^{2} \bar{\lambda}$, $e =g\bar{e}$, and $q = g \bar{q}$.
Note that the mass $m_{\phi}=\sqrt{2\lambda}\eta$ of the Higgs field $\phi_{H}$,
the  mass  $m_{A} = \sqrt{2} e \eta$  of  the  gauge  field  $A^{\mu}$, and the
parameter $\epsilon$  are  invariant  under  the rescaling,  whereas  the  mass
$m_{\chi}$ of the complex scalar field $\chi$ is not subjected to the rescaling.

In terms  of  the  rescaled  fields  and  coupling  constants, self-interaction
potentials (\ref{II:3a}) and (\ref{II:3b}) are written as:
\begin{eqnarray}
U\left( \left\vert \chi \right\vert \right)  &=&\frac{1}{g^{2}}\frac{m_{\chi
}^{2}}{1+\epsilon ^{2}}\left\vert \bar{\chi }\right\vert ^{2}\left[
\left( 1-\left\vert \bar{\chi }\right\vert^{2}\right)^{2}+\epsilon
^{2}\right],                                                        \label{A:3}
\end{eqnarray}
\begin{eqnarray}
V\left( \left\vert \phi \right\vert \right)&=&\frac{1}{g^{2}}
\frac{m_{\phi}^{2}\bar{\eta }^{2}}{4}\left(\frac{\left\vert \bar{\phi}
\right\vert ^{2}}{\bar{\eta }^{2}} -1\right)^{2}.                   \label{A:4}
\end{eqnarray}
Using Eqs.~(\ref{A:3}) and (\ref{A:4}),  it  can  be shown  that the Lagrangian
(\ref{II:1}) has the following behavior under the rescaling:
\begin{gather}
\mathcal{L}\left( \phi, \chi, A^{\mu}, m_{\phi}, \eta, m_{\chi}, g, \epsilon,
e, q \right) = \nonumber \\
g^{-2}\bar{\mathcal{L}}\left( \bar{\phi},\bar{\chi},\bar{A}^{\mu}, m_{\phi},
\bar{\eta}, m_{\chi}, \epsilon, \bar{e}, \bar{q} \right),           \label{A:5}
\end{gather}
where  rescaled  Lagrangian  $\bar{\mathcal{L}} \left(  \bar{\phi}, \bar{\chi},
\bar{A}^{\mu}, m_{\phi}, \bar{\eta}, m_{\chi},\epsilon,\bar{e},\bar{q} \right)$
is $\mathcal{L} \left(\bar{\phi},\bar{\chi},\bar{A}^{\mu}, m_{\phi},\bar{\eta},
m_{\chi}, 1, \epsilon, \bar{e}, \bar{q} \right)$  and  does  not  depend on the
scale factor $g$.
From  Eq.~(\ref{A:5})  it  follows  that if $\left(\phi, \chi, A^{\mu} \right)$
is  a   solution   corresponding  to  the   parameters  $m_{\chi}$, $g_{\chi}$,
$h_{\chi}$,  $\eta$,  $\lambda$,  $e$,  and  $q$  then  $\left(\kappa^{-1}\phi,
\kappa^{-1}\chi, \kappa^{-1} A^{\mu}\right)$  is  also a solution corresponding
to  the  parameters   $m_{\chi}$,  $\kappa^{2}g_{\chi}$,  $\kappa^{4}h_{\chi}$,
$\kappa^{-1}\eta$,   $\kappa^{2}\lambda$,  $\kappa e$,  and  $\kappa q$,  where
$\kappa$ is an arbitrary positive constant.
Thus, if  we  know  a  particular  soliton   solution,  we,  in  fact, know the
one-parameter family of rescaled soliton   solutions.

We want  to  find  the  area  of  the model's  parameters  that  correspond  to
the quasi-classical limit of the soliton system.
To do  this,  we  suppose  that  the  dimensionless  combinations of parameters
of  the  rescaled  Lagrangian  $\bar{\mathcal{L}}$  are  of the order of unity:
\begin{equation}
\epsilon \sim \frac{\bar{e}}{m_{\chi }}\sim \frac{\bar{q}}
{m_{\chi }}\sim \frac{m_{\phi }}{m_{\chi }}\sim  1.                 \label{A:6}
\end{equation}
In this  case,  the  rescaled  soliton  solution $\left(\bar{\phi}, \bar{\chi},
\bar{A}^{\mu}\right)$ is also of the order of unity, whereas the action  of the
soliton system over the period $T= 2 \pi / \omega$ is of the order of $g^{-2}$:
$S_{T}   =   g^{-2}   \int   \nolimits_{0}^{2 \pi /  \omega}  \!\!\! \int dt dx
\bar{\mathcal{L}} \sim g^{-2}$.
Next, we  suppose  that  the  dimensionless  scale  factor  $g$  tends to zero:
\begin{equation}
g = 2\left[ h_{ \chi }/\left( 3g_{ \chi }\right) \right] ^{1/2}  \rightarrow 0.
                                                                    \label{A:7}
\end{equation}
Conditions (\ref{A:6}) and (\ref{A:7}) can be rewritten in terms of the initial
model's parameters as follows:
\begin{eqnarray}
g_{\chi } &\sim &\lambda \sim g^{2}m_{\chi }^{2}\rightarrow 0, \nonumber \\
h_{\chi } &\sim &g^{4}m_{\chi }^{2}\rightarrow 0,              \nonumber \\
e &\sim &q\sim gm_{\chi }\rightarrow 0,                             \label{A:8}
\end{eqnarray}
where the tilde means the same order of magnitude.

Under conditions (\ref{A:8}), the action $S_{T}\left(\phi, \chi, A^{\mu} \right)
=  g^{-2}  \bar{S}_{T}\left( \bar{\phi},  \bar{\chi}, \bar{A}^{\mu} \right)$ is
proportional  to  the  large  factor  $g^{-2}$,  so the  exponential  integrand
$\exp\!\left[i g^{-2} S_{T}\left( \bar{\phi}, \bar{\chi}, \bar{A}^{\mu} \right)
\right]$ of   functional  integrals of quantum field theory  quickly oscillates
as  the  rescaled  fields  $\bar{\phi}$,  $\bar{\chi}$,   and   $\bar{A}^{\mu}$
fluctuate in a neighborhood  of  the  classical  soliton solution.
Due to this, the  main  contribution  to   functional  integrals comes from the
close functional  neighborhood  of  the  classical  soliton solution, so we may
expand the action  $S_{T} \left( \bar{\phi}, \bar{\chi}, \bar{A}^{\mu} \right)$
into the functional series in field variations and retain only  quadratic terms
(linear terms of the functional series vanish due to field equations).
According  to   \cite{Rajaraman},   such   a   behavior   corresponds   to  the
quasi-classical limit of the soliton system.
Thus, the  kink-Q-ball  system  transits  into  the  quasi-classical  limit  if
conditions (\ref{A:8}) are satisfied.
From Eqs.~(\ref{A:8}) it follows that the quasi-classical  limit corresponds to
small  coupling  constants  $h_{\chi }$,  $g_{\chi }$, $\lambda$, $e$, and $q$,
although the coupling  constants differently tend to zero as $g \rightarrow 0$.
At the same time, from the field rescaling it follows that  the quasi-classical
limit corresponds  to  large  amplitudes  of  the  fields  $\phi$,  $\chi$, and
$A^{\mu}$.

Finally, under  the  rescaling,  the  energy  density  and  the  Noether charge
densities behave as follows:
\begin{eqnarray}
\mathcal{E}\left( \phi ,\chi ,A^{\mu }\right)  &=&g^{-2}\mathcal{E}\left(
\bar{\phi },\bar{\chi },\bar{A}^{\mu }\right) ,                    \label{A:9a}
\\
j_{\phi }^{0}\left( \phi ,A^{\mu },e\right)  &=&g^{-2}j_{\phi }^{0}\left(
\bar{\phi },\bar{A}^{\mu },\bar{e}\right) ,                        \label{A:9b}
\\
j_{\chi }^{0}\left( \chi ,A^{\mu },q\right)  &=&g^{-2}j_{\chi }^{0}\left(
\bar{\chi },\bar{A}^{\mu },\bar{q}\right),                         \label{A:9c}
\end{eqnarray}
where in Eq.~(\ref{A:9a}), we omit the lists of parameters (which  are the same
as in Eq.~(\ref{A:5})) for brevity.
From Eqs.~(\ref{A:9a}) -- (\ref{A:9c})  it  follows  that  the  energy  and the
Noether charges of the kink-Q-ball system  become  large in the quasi-classical
limit.

\bibliographystyle{spphys}

\end{document}